\documentclass[reprint, amsmath,amssymb, aps, prd, floatfix, nofootinbib]{revtex4-2}
\usepackage[normalem]{ulem}
\usepackage{graphicx}
\usepackage{dcolumn}
\usepackage[colorlinks=true,linkcolor=blue,citecolor=blue,urlcolor=blue]{hyperref}
\usepackage{bm}
\usepackage{xcolor}
\usepackage[colorlinks=true,
linkcolor=blue,
citecolor=blue,
urlcolor=blue]{hyperref}
\begin{document}

\title{Axion Polarimetric Experiment (APE)}

\author{Qazal Rokn}
\affiliation{Max--Planck-Institut f\"ur Gravitationsphysik (Albert-Einstein-Institut) and Leibniz Universit\"at Hannover, Hannover, Germany}

\author{Ryan Netrval}
\affiliation{Max--Planck-Institut f\"ur Gravitationsphysik (Albert-Einstein-Institut) and Leibniz Universit\"at Hannover, Hannover, Germany}

\author{Aldo Ejlli}
\email[Corresponding author: ]{aldo.ejlli@aei.mpg.de}
\affiliation{Max--Planck-Institut f\"ur Gravitationsphysik (Albert-Einstein-Institut) and Leibniz Universit\"at Hannover, Hannover, Germany}

\author{Guido Mueller}
\affiliation{Max--Planck-Institut f\"ur Gravitationsphysik (Albert-Einstein-Institut) and Leibniz Universit\"at Hannover, Hannover, Germany}
\affiliation{Department of Physics, University of Florida, Gainesville, USA}

\date{\today}
\begin{abstract}
We present the Axion Polarimetric Experiment (APE), a cavity-enhanced polarimeter
designed to search for ultralight axion and axion-like-particle dark matter through
a time-dependent rotation of the linear polarization of laser light. In cavity-based
schemes, intracavity quarter-wave plates can restore coherent buildup of the
axion-induced orthogonal polarization, but their transmissive loss limits the
achievable finesse. To avoid transmissive intracavity optics, we propose a folded
Fabry--P\'erot cavity that employs dielectric phase-shifting mirrors. At an incidence
angle near \(45^\circ\), these mirrors provide a reflection-phase difference
\(\Delta\phi \equiv \phi_s-\phi_p \approx \pi/2\) between \(s\) and \(p\)
polarizations and therefore act as reflective quarter-wave plates. We present the
coating design, thickness optimization, and measurements of the phase shift and
optical loss of the phase-shifting mirrors. Using a heterodyne polarimetric readout
and an explicitly stated noise model, we derive design-level sensitivity projections
for the axion--photon coupling \(g_{a\gamma\gamma}\). These projections should be
interpreted as target sensitivities for the proposed cavity configuration, since the
full-system birefringence noise and angular-jitter coupling remain to be measured.
\end{abstract}

\maketitle

\section{Scientific background and motivation}

The QCD axion was introduced to explain the absence of observable CP violation in the strong interaction. It arises as a pseudo-Nambu--Goldstone boson associated with the spontaneous breaking of a Peccei--Quinn symmetry and dynamically relaxes the effective QCD $\theta$ parameter toward zero~\cite{PhysRevLett.38.1440,PhysRevLett.40.279}.

More general axion-like particles (ALPs) appear in many extensions of the Standard Model, including supergravity and string compactifications. Although they need not solve the strong-CP problem, they can retain a coupling to electromagnetism and therefore admit similar laboratory signatures~\cite{Svrcek:2006yi,PhysRevD.88.035023}.

Astrophysical and cosmological observations require a nonluminous matter component. Together with the absence of confirmed signals in broad regions of weakly interacting massive-particle parameter space, this has strengthened interest in alternative dark-matter candidates~\cite{osti_1454546,XENON:2018voc}. Ultralight axions and ALPs are particularly compelling because nonthermal production can generate a highly occupied Galactic halo field that is well described locally as a coherent classical oscillation.

Through the axion--photon coupling $g_{a\gamma\gamma}$, such a background can induce a small time-dependent rotation of the polarization of linearly polarized light. In the dark-matter scenario, the signal oscillates at
\[
\omega_a = \frac{m_a c^2}{\hbar},
\]
with a coherence time set by the Galactic velocity distribution. The precise definition of the rotation observable used in this work is given in Sec.~II.

Cavity-enhanced polarimetry provides a direct way to search for this effect by accumulating a small polarization signal over many effective passes. Dedicated cavity-polarimetry efforts include LIDA~\cite{Heinze:2023nfb}, DANCE~\cite{Oshima:2023csb}, and ADBC~\cite{Pandey2024ADBC}. Precision polarimetry experiments such as PVLAS~\cite{PhysRevD.77.032006}, while primarily developed for vacuum magnetic birefringence and related light-boson searches, also demonstrate the sensitivity of high-finesse optical systems to extremely small polarization signals.

Related proposals aim to extend axion polarimetry to large-scale interferometers, including gravitational-wave detectors, by reading out axion-induced polarization signals in auxiliary ports or transmitted beams; KAGRA, for example, has developed and installed polarization optics in this direction~\cite{Michimura:2025kod,Michimura:2021hwr}. More broadly, gravitational-wave interferometer data have also been used to search for ultralight dark matter through apparent differential-strain signals~\cite{Gottel:2024cfj,LVK:2025dm}. Complementary laboratory searches for light bosons include the ALPS~II light-shining-through-a-wall program~\cite{ALPSII:2025eri,Spector:2026eys}.

In this work we introduce the Axion Polarimetric Experiment (APE), a cavity-enhanced polarimeter being developed at the Max Planck Institute for Gravitational Physics. APE is implemented in two stages. In the first stage, the experiment employs a Fabry--P\'erot cavity containing two intracavity quarter-wave plates (QWPs). This configuration provides an end-to-end demonstration of the vacuum polarimetric readout and prevents cancellation of the axion-induced signal over successive cavity round trips by controlling the intracavity polarization evolution~\cite{PhysRevD.107.083035,Rokn:2025oyw}. In the second stage, the transmissive wave plates are replaced by phase-shifting mirrors in a folded-cavity geometry. This removes transmissive intracavity optics, reduces round-trip loss, and enables higher finesse, thereby improving the projected sensitivity to $g_{a\gamma\gamma}$ in the ultralight-mass regime.

We present the design logic of both stages, develop the corresponding noise model, and derive
sensitivity projections for the upgraded phase-shifting-mirror configuration.

\section{Polarization signatures of axion--photon interactions}

We work in axion electrodynamics, described by the Lagrangian density~\cite{Mirizzi:2006zy}
\begin{equation}
\begin{split}
\mathcal{L}
&= -\frac{1}{4}F_{\mu\nu}F^{\mu\nu}
-\frac{1}{4} g_{a\gamma\gamma}\, a(t)\, F_{\mu\nu}\tilde F^{\mu\nu} \\
&= \frac{1}{2}\!\left(\mathbf{E}^2-\mathbf{B}^2\right)
+ g_{a\gamma\gamma}\, a(t)\,\mathbf{E}\!\cdot\!\mathbf{B},
\end{split}
\label{eq:L_axion}
\end{equation}
where \(\tilde F^{\mu\nu}\equiv \tfrac12\epsilon^{\mu\nu\rho\sigma}F_{\rho\sigma}\). We assume
that the axion background is spatially homogeneous over the optical path, \(a=a(t)\), and we
use Heaviside--Lorentz units with \(c=1\).

In the presence of the axion--photon interaction, the constitutive relations become
\begin{align}
\mathbf{D} &= \mathbf{E} + g_{a\gamma\gamma} a(t)\,\mathbf{B},
\\
\mathbf{H} &= \mathbf{B} - g_{a\gamma\gamma} a(t)\,\mathbf{E}.
\end{align}
Equivalently, vacuum behaves as a lossless magneto-electric medium with effective polarization
and magnetization
\begin{align}
\mathbf{P} &= \mathbf{D}-\mathbf{E}
= g_{a\gamma\gamma} a(t)\,\mathbf{B},
\\
\mathbf{M} &= \mathbf{B}-\mathbf{H}
= g_{a\gamma\gamma} a(t)\,\mathbf{E}.
\end{align}
A time-dependent axion background therefore induces circular birefringence.

For a spatially homogeneous axion field, \(\nabla a \simeq 0\), and in the absence of free
charges and currents, Maxwell's equations give
\begin{equation}
\nabla\times\mathbf{B}-\dot{\mathbf{E}}
=
g_{a\gamma\gamma}\,\dot a(t)\,\mathbf{B}.
\label{eq:Ampere_mod}
\end{equation}
Together with \(\nabla\times\mathbf{E}=-\dot{\mathbf{B}}\), this implies that the two circular
polarization eigenmodes propagate differently. For a plane wave traveling along \(+\hat z\), it
is convenient to introduce the circular basis
\begin{equation}
\mathbf{e}_\pm \equiv \frac{1}{\sqrt2}\,(\hat x \pm i \hat y),
\end{equation}
which diagonalizes the action of \(\hat z\times\). In the adiabatic limit
\(\omega_a\ll\omega\), and to first order in \(g_{a\gamma\gamma}\dot a/\omega\ll1\), the two
circular eigenmodes acquire different wavenumbers,
\begin{equation}
k_\pm \simeq \omega \mp \frac{g_{a\gamma\gamma}}{2}\,\dot a(t),
\label{eq:kpm}
\end{equation}
so the axion background produces circular birefringence, but no dichroism at this order.

The relative phase accumulated between the two circular components over a one-way propagation
time \(L\) rotates the linear polarization by \(\beta=\Delta\phi/2\), yielding~\cite{PhysRevD.107.083035}
\begin{equation}
\beta(t,L)=\frac{g_{a\gamma\gamma}}{2}\bigl[a(t)-a(t-L)\bigr].
\label{eq:beta_general}
\end{equation}
For virialized axion dark matter, the field is locally well approximated over a coherence time
by a classical oscillation \(a(t)=a_0\cos(\omega_a t+\varphi_a)\). Choosing the phase origin so
that \(\varphi_a=0\), one finds
\begin{equation}
\beta(t,L)=
-\,g_{a\gamma\gamma}a_0
\sin\!\left(\frac{\omega_a L}{2}\right)
\sin\!\left[\omega_a\!\left(t-\frac{L}{2}\right)\right].
\label{eq:beta_sineform}
\end{equation}
In the short-baseline limit \(\omega_a L\ll1\),
\begin{equation}
\beta(t,L)\simeq
\frac{g_{a\gamma\gamma}}{2}\,L\,\dot a\!\left(t-\frac{L}{2}\right),
\label{eq:beta_small_omegaL}
\end{equation}
which makes explicit that the effect is driven by the time variation of the axion field.
In the small-angle limit, the axion interaction generates a field amplitude in the polarization channel orthogonal to the carrier. Writing
\begin{equation}
\mathbf{E}_{\parallel}(t)=E_{\parallel}(t)\,\hat{\mathbf e}_{\parallel},
\qquad
\mathbf{E}_{\rm a}(t,L)=E_{\rm a}(t,L)\,\hat{\mathbf e}_{\perp},
\end{equation}
with \(\hat{\mathbf e}_{\perp}\cdot\hat{\mathbf e}_{\parallel}=0\), the one-pass axion-induced scalar field amplitude is
\begin{equation}
E_{\rm a}(t,L)\simeq \beta(t,L)\,E_{\parallel}(t-L),
\label{eq:Ea_def}
\end{equation}
where \(\{\hat{\mathbf e}_{\parallel},\hat{\mathbf e}_{\perp}\}\) denotes the linear laboratory polarization basis used for the cavity readout, while the circular basis introduced above diagonalizes propagation in the axion background. Here \(\beta(t,L)\ll1\) is the axion-induced polarization rotation accumulated over the propagation length \(L\). In the regime \(\omega_aL\ll1\), \(\beta(t,L)\) varies negligibly over a cavity round trip, providing the basis for the cavity-enhancement analysis developed in the next section.

\section{Axion-Induced Polarization Enhancement Using QWPs in a Fabry--P\'erot Cavity}
\label{sec:qwp-section}
In this experiment, a purely p-polarized optical field is injected into the system. A hypothetical axion-induced interaction would produce a small rotation of the polarization state, thereby generating an orthogonal s-polarized field component. The amplitude of this s-polarized signal is directly related to the axion-induced polarization rotation angle.

\begin{figure}[tbp]
\centering
\includegraphics[width=0.9\linewidth]{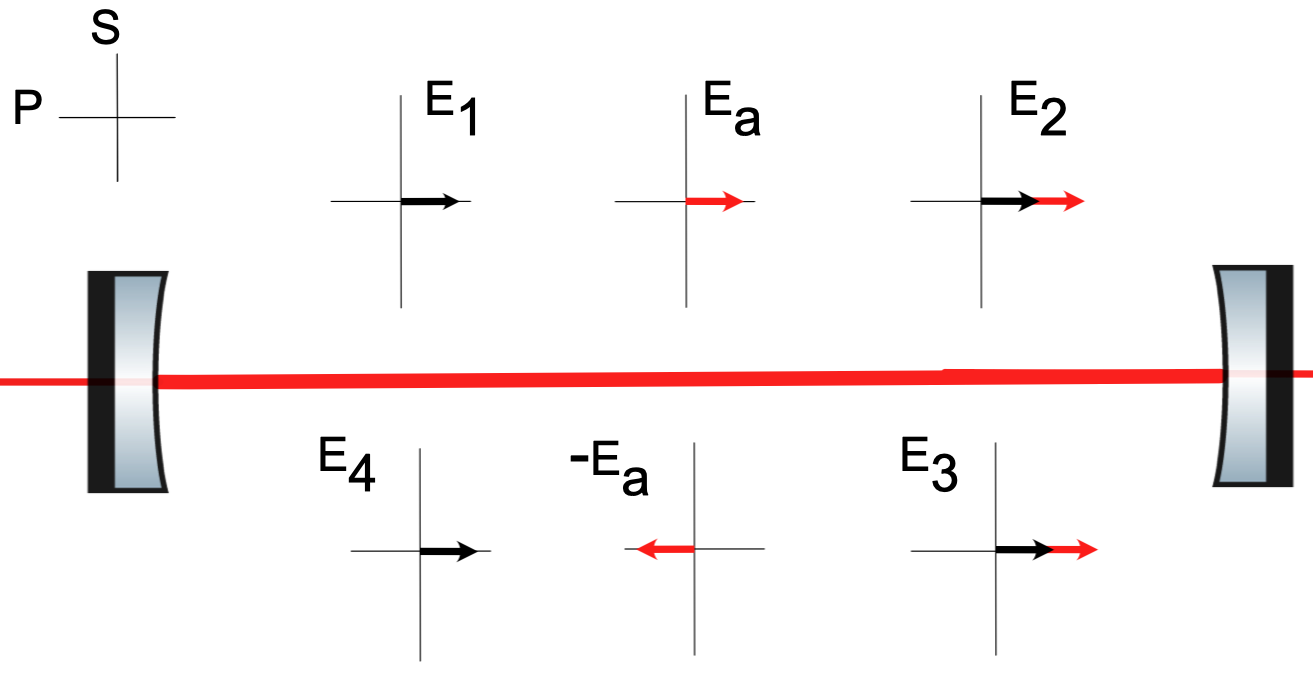}
\caption{Schematic of the orthogonal signal field in an empty linear cavity. The vectors
\(\mathbf{E}_{1\ldots4}\) denote the orthogonal field component at the cavity mirrors in a fixed
laboratory \(s/p\) basis. Over the forward half trip, the axion interaction generates an
increment \(+\mathbf{E}_{\rm a}\), whereas over the reverse half trip it generates
\(-\mathbf{E}_{\rm a}\). The two contributions therefore cancel over one round trip, so the
signal field does not acquire the usual resonant buildup.}
\label{fig:empty-cavity}
\end{figure}

In an empty linear cavity, the axion-induced orthogonal field generated on successive half
trips enters with opposite sign in a fixed laboratory polarization basis. As a result, the
signal does not build up resonantly. Figure~\ref{fig:empty-cavity} illustrates this field-label
convention and the sign reversal of the generated orthogonal component under propagation
reversal.

Eliminating the intermediate fields then gives the steady-state orthogonal signal field at the
reference plane,
\begin{equation}
\mathbf{E}_2 \simeq \frac{1-r}{1-r^2}\,\mathbf{E}_{\rm a}
= \frac{1}{1+r}\,\mathbf{E}_{\rm a},
\label{eq:E2_empty_ss}
\end{equation}

where \(r\) is the real positive field-amplitude reflectivity of each cavity mirror. Thus,
\(\mathbf{E}_2 \to \mathbf{E}_{\rm a}/2\) as \(r\to 1\): the axion-induced orthogonal field
remains of order unity and does not acquire the usual resonant factor \((1-r)^{-1}\).

\subsection*{With two QWPs: restoring coherent addition}

\begin{figure}[!tbp]
    \centering
    \includegraphics[width=0.9\linewidth]{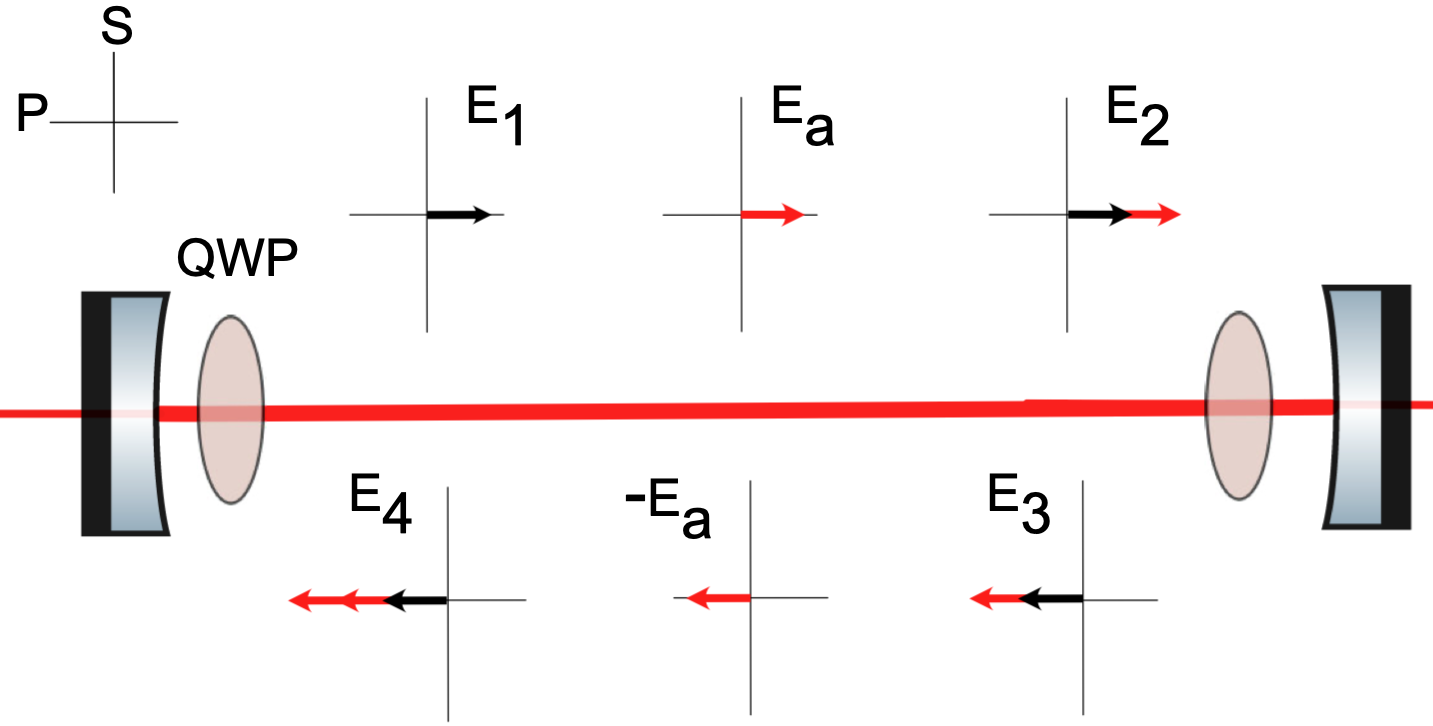}
    \caption{Linear cavity with two intracavity QWPs. The QWPs are oriented so that the carrier
    polarization is restored after each round trip, while the axion-induced orthogonal field
    generated on the return half trip acquires the sign required to add constructively with the
    forward-half-trip contribution in a fixed laboratory \(s/p\) basis. The destructive
    cancellation of the empty cavity is thereby converted into constructive round-trip addition,
    restoring resonant buildup of the signal field.}
    \label{fig:qwp-cavity}
\end{figure}

To prevent cancellation of the orthogonal signal generated on successive half trips, one may
insert two QWPs inside the cavity. The QWPs are oriented such that the
carrier polarization is restored after a full round trip, while the orthogonal field generated
on the return half trip enters the fixed laboratory basis with the sign required for
constructive addition. In the signal-channel recursion, this changes only the return-half-trip
relation:
\begin{equation}
\mathbf{E}_4(t)=\mathbf{E}_3(t-L)+\mathbf{E}_{\rm a}(t,L).
\label{eq:qwp_E4}
\end{equation}
All other relations remain identical to the empty-cavity case.

Repeating the same elimination as above yields
\begin{equation}
\mathbf{E}_2(t)\simeq r^2\,\mathbf{E}_2(t-2L)+(1+r)\,\mathbf{E}_{\rm a}(t,L),
\label{eq:E2_qwp_clean}
\end{equation}
so that the axion-induced contributions from successive round trips add constructively. In the
regime where the signal varies negligibly over one round trip, i.e.\ for \(\omega_a L\ll1\),
the steady-state solution becomes
\begin{equation}
\mathbf{E}_2 \simeq \frac{1+r}{1-r^2}\,\mathbf{E}_{\rm a}
= \frac{1}{1-r}\,\mathbf{E}_{\rm a}.
\label{eq:E2_qwp_ss}
\end{equation}
The two-QWP configuration therefore restores the usual resonant \emph{field} buildup of the
axion-induced orthogonal polarization.

Although the QWP-based configuration solves the cancellation problem, transmissive intracavity
optics introduce additional loss, which limits the achievable finesse and hence the ultimate
sensitivity (see Sec.~\ref{section-sensitivity}). This motivates the reflective configuration
introduced in the next section, in which the QWPs are replaced by phase-shifting mirrors to
reduce intracavity loss while preserving coherent addition of the axion-induced signal.

\section{Alternative cavity design: replacing QWPs with phase-shifting mirrors}
\label{sec:psm-cavity}

The QWP-based scheme of Sec.~III enables coherent buildup of the axion-induced
orthogonal field, but transmissive intracavity optics introduce additional loss. For the
wave plates used in our reference design, the vendor coating inspection report gives
residual AR-coating reflectances at 1064~nm of \(R=0.0025\%\) on surface S1 and
\(R=0.0346\%\) on surface S2, corresponding to approximately
\(25~\mathrm{ppm}\) and \(3.5\times10^2~\mathrm{ppm}\), respectively.\footnote{CASTECH
P/N 040003, batch P86030-1, PO 2025-21142, vendor coating inspection report dated
18 June 2025.} We treat the sum of these two surface reflectances as an effective
single-traversal insertion loss, \(l_{\rm QWP,pass}\simeq371~\mathrm{ppm}\). Since
the linear cavity contains two QWPs and each is traversed twice per round trip, the
corresponding round-trip loss contribution is
\(l_{\rm int,rt}^{\rm QWP}\simeq4l_{\rm QWP,pass}\simeq1.48\times10^3~\mathrm{ppm}\).
These losses directly reduce the achievable finesse and therefore the projected
sensitivity.

\begin{figure}[!tbp]
\centering
\includegraphics[width=0.75\linewidth]{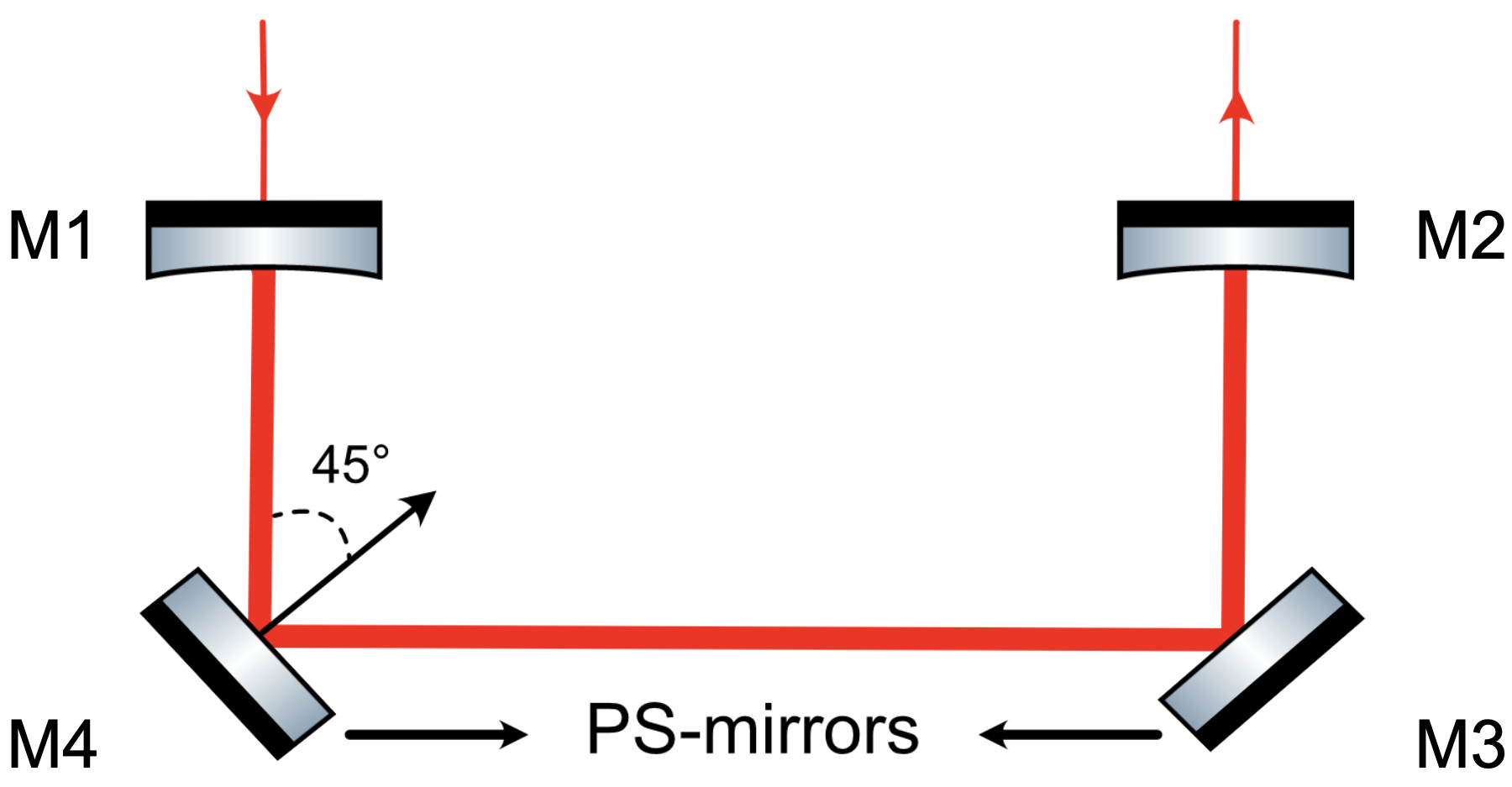}
\caption{Folded Fabry--P\'erot cavity employing two phase-shifting mirrors (PSMs),
M3 and M4, operated at an incidence angle near \(45^\circ\). The end mirrors M1
and M2 are close to normal incidence, while the two folding mirrors provide the
polarization-dependent reflection phase that replaces the function of the intracavity
QWPs.}
\label{fig:psm_cavity}
\end{figure}

To reduce intracavity loss while preserving coherent addition of the axion-induced
signal, we replace the transmissive QWPs by two reflective phase-shifting mirrors in
a folded Fabry--P\'erot geometry, as shown in Fig.~\ref{fig:psm_cavity}. The end
mirrors are close to normal incidence, while the two folding mirrors are operated at
an incidence angle near \(45^\circ\).

For a reflection at incidence angle \(\theta\), the mirror coating imparts
polarization-dependent complex reflection coefficients,
\begin{align}
E_s &\to r_s(\theta)e^{i\phi_s(\theta)}E_s,\\
E_p &\to r_p(\theta)e^{i\phi_p(\theta)}E_p,
\end{align}
where \(s\) and \(p\) denote the polarization components defined with respect to the
local plane of incidence at each mirror. We define the single-reflection differential
phase
\begin{equation}
\Delta\phi(\theta)\equiv \phi_s(\theta)-\phi_p(\theta),
\label{eq:psm_delta_phi}
\end{equation}
which acts as a reflective wave-plate retardance in the local \(s/p\) basis. For the
coating design considered here, the mirrors are optimized near
\(\theta_0\simeq45^\circ\) such that
\begin{equation}
\Delta\phi(\theta_0)\approx\frac{\pi}{2},
\label{eq:psm_quarter_wave_condition}
\end{equation}
while maintaining high reflectance for both polarizations.

\paragraph*{Jones-map condition for coherent addition.}

The equivalence between the phase-shifting-mirror cavity and the two-QWP
configuration can be stated explicitly in Jones form. After removing a common
reflection phase, the Jones matrix of the \(i\)-th phase-shifting mirror in its
local \((p,s)\) basis may be written as
\begin{equation}
J_i^{\rm loc}
=
\begin{pmatrix}
1 & 0\\
0 & \eta_i e^{i\Delta\phi_i}
\end{pmatrix},
\qquad
\eta_i\equiv\frac{r_{s,i}}{r_{p,i}},
\label{eq:psm_jones_local}
\end{equation}
where \(\Delta\phi_i=\phi_{s,i}-\phi_{p,i}\). In the ideal low-loss limit,
\(\eta_i\simeq1\). A half trip of the folded cavity contains two PSM reflections.
If the two local incidence bases are aligned, or if the corresponding
basis-transport rotations are absorbed into the definition of the laboratory
\(p/s\) basis, the half-trip polarization map is
\begin{equation}
J_{\rm half}
=
J_4^{\rm loc}J_3^{\rm loc}
\simeq
\begin{pmatrix}
1 & 0\\
0 & e^{i(\Delta\phi_3+\Delta\phi_4)}
\end{pmatrix}.
\label{eq:psm_half_trip_map}
\end{equation}
For \(\Delta\phi_3\simeq\Delta\phi_4\simeq\pi/2\), this becomes
\begin{equation}
J_{\rm half}
\simeq
\begin{pmatrix}
1 & 0\\
0 & -1
\end{pmatrix}.
\label{eq:psm_half_trip_pi}
\end{equation}
Thus the carrier \(p\)-polarization is preserved, while the orthogonal
\(s\)-polarized component acquires an additional sign change over one half trip.
This sign change maps the axion-induced field generated on the return half trip
into the same sign convention as the field generated on the forward half trip, so
the two source terms add constructively rather than canceling as in an empty
linear cavity. Over a full round trip,
\begin{equation}
J_{\rm rt}=J_{\rm half}^2\simeq \mathbf I,
\label{eq:psm_round_trip_identity}
\end{equation}
up to a common optical phase. The carrier therefore remains an eigenpolarization
of the cavity, while the orthogonal signal field has the round-trip sign
convention needed for resonant buildup.

If the local planes of incidence are not aligned, the local Jones matrix must be
rotated into a common laboratory basis,
\begin{align}
J_i(\alpha_i)
&=
R(-\alpha_i)J_i^{\rm loc}R(\alpha_i),
\\
R(\alpha)
&=
\begin{pmatrix}
\cos\alpha & -\sin\alpha\\
\sin\alpha & \cos\alpha
\end{pmatrix}.
\end{align}
The full folded-cavity map is then obtained schematically from the ordered product
\[
J_{\rm rt}
=
\mathcal P\prod_{\ell\in{\rm rt}}J_\ell(\alpha_\ell),
\]
where \(\mathcal P\) denotes ordering along the propagation path. For the
sensitivity projections we assume that the PSM pair is aligned such that the
\(p\)-polarized carrier is an eigenpolarization of this full round-trip map. The
eigenvalues of \(J_{\rm rt}\) determine the polarization phase
\(\Phi_{\rm pol}\) used in the signal-mode transfer function of
Sec.~\ref{section-sensitivity}. A full numerical evaluation with the final cavity
geometry will be required to determine the exact eigenpolarizations and
polarization-mode splitting of the assembled cavity.

A practical advantage of the reflective implementation is that it avoids
transmissive intracavity optics. The dominant losses are then coating
transmission, absorption, and scatter, which can in principle be made smaller than
the losses of transmissive wave plates at comparable optical quality. The PSM
coatings are realized as alternating high- and low-index dielectric layers
optimized to satisfy simultaneously: (i) high power reflectance for both
polarizations,
\begin{equation}
R_s(\theta_0),\,R_p(\theta_0)>0.99995,
\end{equation}
and (ii) a differential phase
\begin{equation}
\Delta\phi(\theta_0)\approx\frac{\pi}{2}
\end{equation}
at \(\theta_0\simeq45^\circ\). The multilayer design and thickness optimization
are described in Sec.~\ref{sec:thickness-optimization} and
Appendix~\ref{app:coating}.

In the sensitivity estimate of Sec.~\ref{section-sensitivity}, we use the
measured polarization-dependent transmission together with the scatter and
absorption budget to compute the achievable finesse and the resulting improvement
relative to the QWP configuration, rather than assuming an \emph{a priori} gain.

\section{Thickness optimization for a \texorpdfstring{$\boldsymbol{\pi/2}$}{π/2} phase-shifting dielectric coating}
\label{sec:thickness-optimization}

The phase-shifting-mirror coating was optimized to satisfy two requirements at the operating
incidence angle $\theta_0$: a differential reflection phase
\(
\Delta\phi \equiv \phi_s-\phi_p \approx \pi/2
\)
and high reflectance for both $s$ and $p$ polarizations. Appendix~\ref{app:coating}
summarizes the transfer-matrix formalism used in the coating calculation; here we state the
optimization target, the resulting angular response, and its robustness to layer-thickness
errors.

As the starting point of the numerical search, each layer thickness was chosen near the
quarter-wave optical-thickness condition at $(\lambda_0,\theta_0)$, and the optimizer was then
allowed to introduce controlled deviations to satisfy both the phase and reflectance targets.
For a candidate layer-thickness vector $\{d_j\}$, we minimize the merit function
\begin{equation}
\begin{aligned}
\mathcal{J}(\{d_j\}) &= \delta_\phi^2
+ \alpha\left(\Delta R_s^2+\Delta R_p^2\right),\\
\Delta R_j &\equiv \max\!\left(0,\,R_{\rm target}-R_j\right),
\qquad (j=s,p),
\end{aligned}
\label{eq:cost_function}
\end{equation}
evaluated at the design angle $\theta_0$, where $R_{\rm target}$ is the minimum allowed
reflectance and $\alpha$ weights the reflectance penalty relative to the phase target. The
phase error is computed with the wrapped difference
\begin{equation}
\delta_\phi \equiv
\mathrm{Arg}\!\left[\exp\!\big(i(\Delta\phi(\theta_0)-\pi/2)\big)\right]\in(-\pi,\pi],
\label{eq:wrapped_phase}
\end{equation}
so that the optimization drives $\Delta\phi(\theta_0)$ toward $\pi/2$ independent of the
$2\pi$ branch choice.

\begin{figure}[!tbp]
  \centering
  \includegraphics[width=0.95\linewidth]{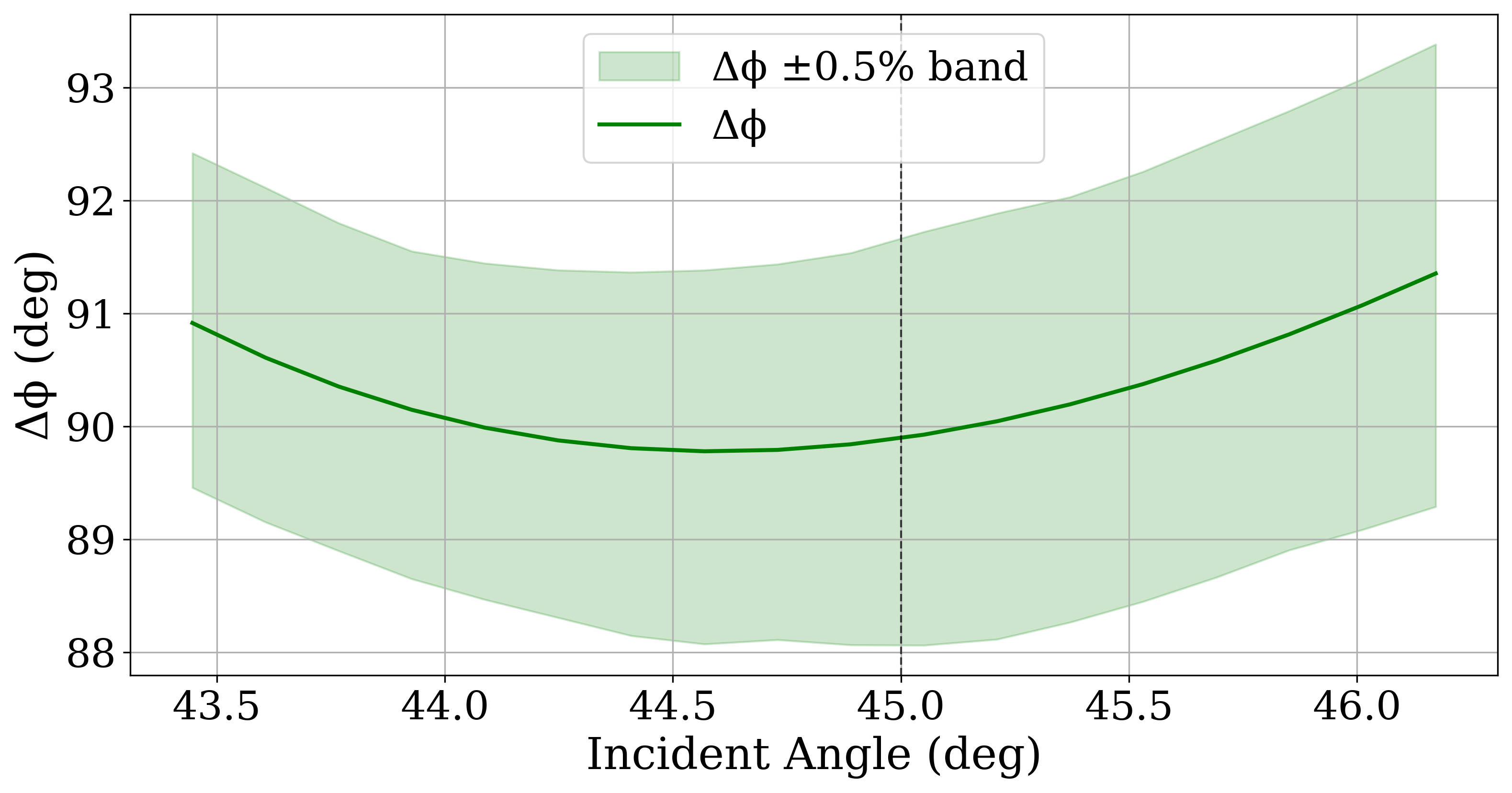}
  \caption{Calculated differential phase $\Delta\phi=\phi_s-\phi_p$ versus incidence angle near
  the design angle $\theta_0\approx45^\circ$. Solid curve: nominal optimized design. Shaded
  band: central 95\% interval from 200 realizations with independent $\pm0.5\%$ random
  perturbations of each layer thickness.}
  \label{fig:phase_simulation}
\end{figure}

Figure~\ref{fig:phase_simulation} shows the calculated differential phase
$\Delta\phi(\theta)$ for the nominal optimized coating. Near the design angle
$\theta_0 \approx 45^\circ$, the coating provides the target quarter-wave retardance,
$\Delta\phi \approx 90^\circ$. To estimate fabrication tolerance, we performed 200 Monte Carlo
realizations in which each layer thickness was independently perturbed by up to $\pm0.5\%$. The
resulting central 95\% band shows that the main effect of thickness errors is a modest shift in
the angle at which $\Delta\phi \approx 90^\circ$ is reached, while the overall angular
dependence remains similar to the nominal design.

\begin{figure}[!tbp]
\centering
\includegraphics[width=0.95\linewidth]{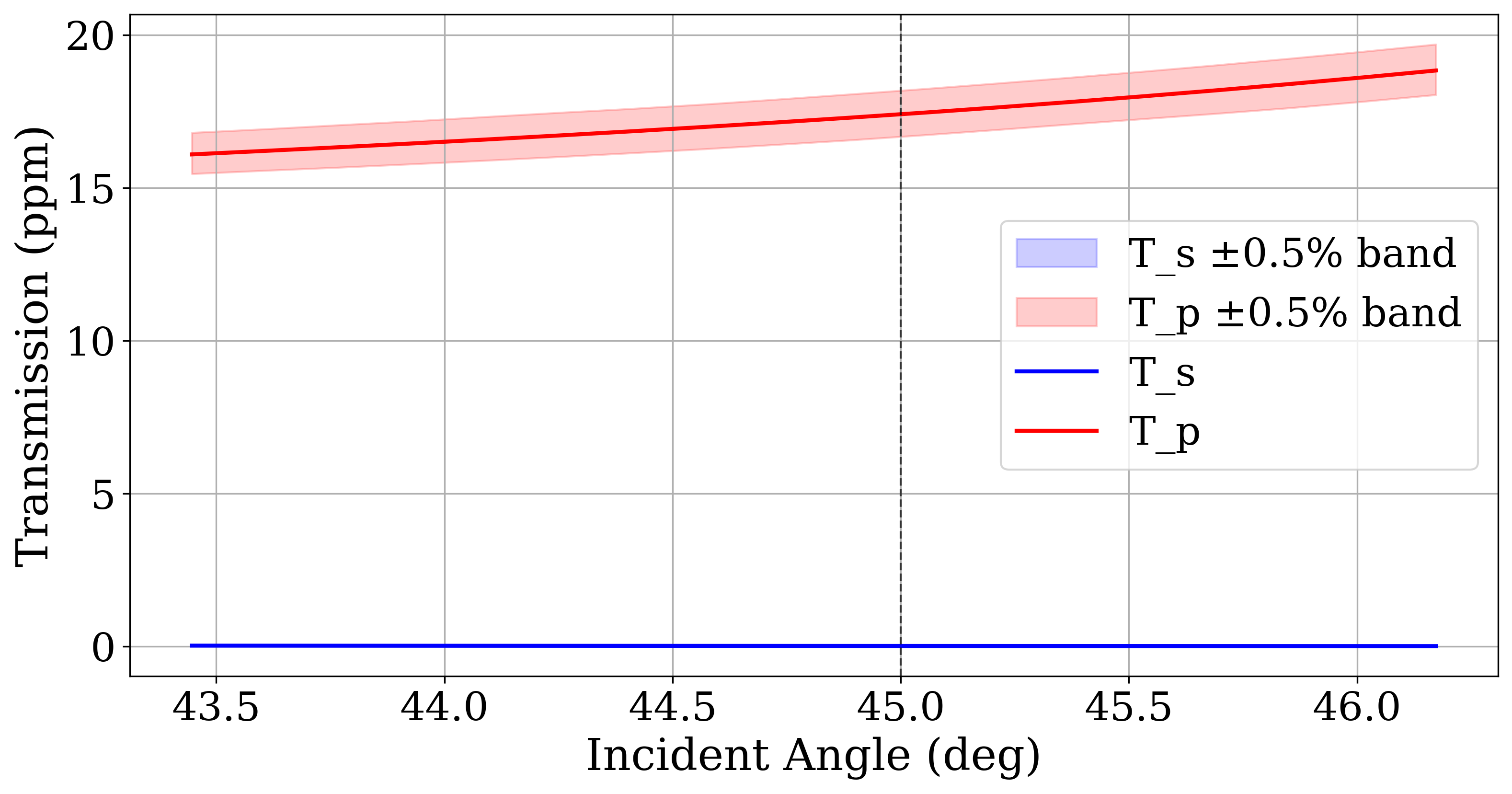}
\caption{Calculated $s$- and $p$-polarized residual transmissions versus incidence angle near
the design angle $\theta_0\approx45^\circ$. Solid curves: nominal optimized design. Shaded
band: central 95\% interval from 200 realizations with independent $\pm0.5\%$ thickness
perturbations. Near $\theta_0$, both polarizations remain in the low-loss regime.}
\label{fig:transmission_simulation}
\end{figure}

Figure~\ref{fig:transmission_simulation} shows the corresponding $s$- and $p$-polarized
residual transmissions for the same ensemble. Around $\theta_0$, the residual transmission
remains small for both polarizations, and the spread induced by the thickness perturbations is
limited. Thus, in the target angular range, the optimized coating preserves both the required
differential phase and the low optical loss needed for high-finesse cavity operation.

After optimization, the coating was fabricated. The vendor introduced minor adjustments to the
final layer-thickness set to satisfy manufacturability constraints, such as thickness rounding
within the specified tolerances. In the next section, we compare the measured differential phase
of the fabricated mirrors with the transfer-matrix simulations.

\section{Experimental characterization}
\label{sec:characterization}

The fabricated phase-shifting mirrors were characterized using the ellipsometric setup
described in Appendix~\ref{app:ellipsometry}. Throughout this work, the reported
phase is the differential reflection phase in the local incidence basis,
\begin{equation}
\Delta\phi \equiv \phi_s-\phi_p,
\end{equation}
consistent with the convention adopted in Sec.~\ref{sec:psm-cavity}.

\begin{figure}[t]
  \centering
  \includegraphics[width=0.95\linewidth]{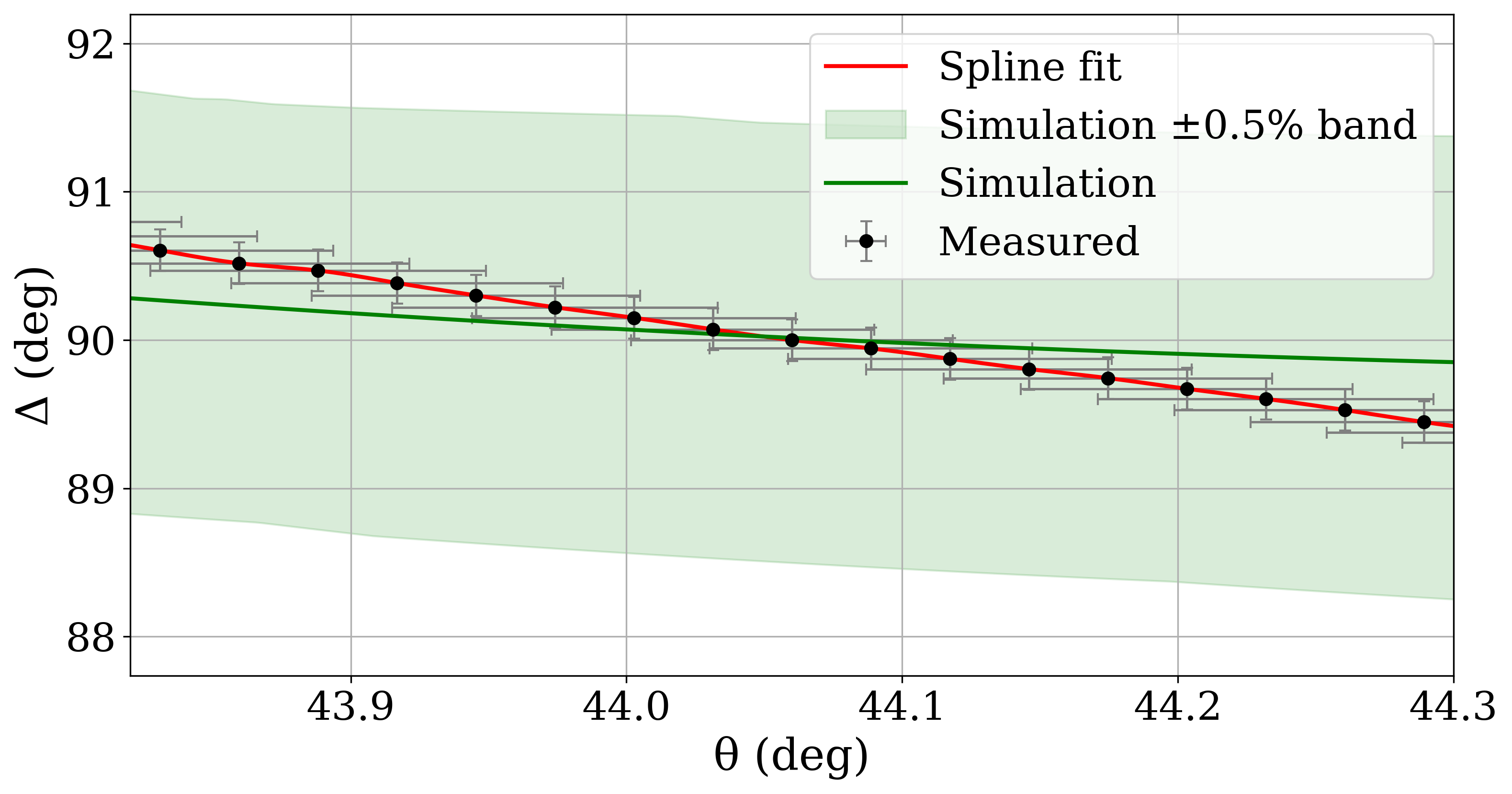}
  \caption{Measured differential phase $\Delta\phi$ versus incidence angle for a fabricated
phase-shifting mirror, compared with transfer-matrix simulations. The mirror reaches
$\Delta\phi\simeq90^\circ$ near $\theta\simeq44^\circ$. The measured curve falls within
the simulated tolerance band corresponding to independent $\pm0.5\%$ thickness perturbations
of each layer, shown in Fig.~\ref{fig:phase_simulation}.}
  \label{fig:characterization}
\end{figure}

Figure~\ref{fig:characterization} compares the measured angular dependence
of \(\Delta\phi(\theta)\) with the transfer-matrix prediction for the fabricated mirror.
The differential phase reaches \(\Delta\phi\simeq90^\circ\) at an incidence angle close
to the design value. The measured response lies within the simulated tolerance band
shown in Fig.~\ref{fig:phase_simulation}, indicating that the fabricated coating is
consistent with the nominal multilayer design and the assumed fabrication tolerances.

We also measured the polarization-dependent power transmission at the operating angle. At
\(\theta \simeq 44^\circ\) and \(\lambda_0=1064~\mathrm{nm}\), the single-reflection
transmissions of one phase-shifting mirror were
\begin{equation}
T_s^{\rm PSM} \simeq 0.1~\mathrm{ppm},
\qquad
T_p^{\rm PSM} \simeq 19~\mathrm{ppm},
\end{equation}
for the local \(s\) and \(p\) components, respectively. Since the folded cavity contains
two phase-shifting mirrors and each is encountered twice per round trip, there are four
PSM reflections per cavity round trip.

The transmission contribution to the round-trip internal loss is therefore
\begin{equation}
\begin{aligned}
l^{\rm PSM,T}_{{\rm rt},s}
&\simeq 4T_s^{\rm PSM}
\simeq 0.4~{\rm ppm},\\
l^{\rm PSM,T}_{{\rm rt},p}
&\simeq 4T_p^{\rm PSM}
\simeq 76~{\rm ppm}.
\end{aligned}
\end{equation}
Dedicated scatter and absorption measurements of the fabricated coated mirrors have not
yet been performed. For the design-level loss budget we therefore assign a conservative
non-transmission loss \(l_{\rm SA}^{\rm PSM}=2~\mathrm{ppm}\) per PSM reflection from
coating scatter and absorption. The total PSM round-trip losses used in
Table~\ref{tab:cavity_specs} are then
\begin{equation}
\begin{aligned}
l^{\rm PSM}_{{\rm int,rt},s}
&\simeq 4\left(T_s^{\rm PSM}+l_{\rm SA}^{\rm PSM}\right)
\simeq 8.4~{\rm ppm},\\
l^{\rm PSM}_{{\rm int,rt},p}
&\simeq 4\left(T_p^{\rm PSM}+l_{\rm SA}^{\rm PSM}\right)
\simeq 84~{\rm ppm}.
\end{aligned}
\end{equation}
These values should be interpreted as design-level loss estimates pending a direct
cavity linewidth or ringdown measurement with the installed PSMs.

These measurements show that the fabricated mirror reproduces the target phase response near the
operating angle and exhibits the expected polarization-dependent loss. In Sec.~\ref{section-sensitivity},
we use the measured transmission together with the assumed non-transmission losses to quantify
the corresponding impact on the projected sensitivity to the axion--photon coupling.

\section{Sensitivity to the axion--photon coupling}
\label{section-sensitivity}

We derive the projected sensitivity of the cavity polarimeter to the axion--photon
coupling \(g_{a\gamma\gamma}\). The \(p\)-polarized carrier at optical frequency
\(\omega_0\) is treated as the pump field and is held on resonance by the cavity
locking system. The axion field generates a weak orthogonal \((s\)-polarized)
signal at sideband frequencies \(\omega_0\pm\omega_a\). We first derive the
cavity transfer function for the signal field and then relate it to the
corresponding rotation sensitivity.

The sensitivity curves presented below are design-level projections rather than
measured exclusion limits. They combine the cavity response with measured or
estimated optical losses, the photon-counting and heterodyne readout models, and
the empirical coating-birefringence benchmark introduced in
Sec.~\ref{subsec:open_biref_discussion}. Angular-jitter coupling is not included as a
measured noise contribution, but is treated as a commissioning requirement to be
verified with the assembled cavity. The curves therefore quantify the expected
reach of the proposed optical architecture under the stated assumptions, not the
demonstrated performance of an operating dark-matter detector.

\subsection{Pump and signal buildup}

Let $E_{0,p}$ denote the incident $p$-polarized field amplitude. We define the
round-trip \emph{amplitude} loss factors
\begin{equation}
\rho_p \equiv r_{1p}\,r_{2p}\,r_{3p}^2\,r_{4p}^2,
\qquad
\rho_s \equiv r_{1s}\,r_{2s}\,r_{3s}^2\,r_{4s}^2,
\label{eq:rt_factors_clean}
\end{equation}
where the $r_{ij}$ are taken to be real positive amplitude factors and mirrors~3
and~4 are encountered twice per round trip in the folded geometry. Mirrors~1 and~2
are at normal incidence, so $r_{1s}=r_{1p}\equiv r_1$ and
$r_{2s}=r_{2p}\equiv r_2$. Any polarization-dependent internal loss may be folded
into $\rho_{s,p}$ through additional amplitude factors.

\paragraph{(i) Pump buildup.}
On resonance, the circulating pump field is
\begin{equation}
E_p^{(\mathrm{circ})}
=
\frac{t_1^p}{1-\rho_p}\,E_{0,p},
\label{eq:pump_circ_clean}
\end{equation}
where $t_1^p$ is the input-coupler amplitude transmission. The corresponding pump
field buildup factor is
\begin{equation}
G_p \equiv \frac{E_p^{(\mathrm{circ})}}{E_{0,p}}
=
\frac{t_1^p}{1-\rho_p}.
\label{eq:Gp_def_clean}
\end{equation}

\paragraph{(ii) Signal buildup.}
The earlier cavity-polarization analysis reduces the axion effect to an effective
one-pass source term in the orthogonal polarization. Let
$\beta_0(\omega_a,L)$ denote the \emph{peak} single-pass rotation amplitude,
\begin{equation}
\beta_0(\omega_a,L)
=
g_{a\gamma\gamma}\,a_0\,\bigl|\sin(\omega_a L/2)\bigr|,
\label{eq:beta0_def}
\end{equation}
where $a_0$ is fixed by the local axion energy density. If one instead works with
the complex envelope of a single sideband, its magnitude is $\beta_0/2$; below we
keep the real peak amplitude convention explicit to avoid factor-of-2 ambiguity.

The orthogonal field generated in one pass is then
\begin{equation}
E_s^{(0)}(\omega_a)=\beta_0(\omega_a,L)\,E_p^{(\mathrm{circ})},
\label{eq:Es0_clean}
\end{equation}
up to an irrelevant phase convention for the signal envelope. The circulating
signal field at offset frequency $\omega_a$ is
\begin{equation}
E_s^{(\mathrm{circ})}(\omega_a)=
\frac{E_s^{(0)}(\omega_a)}
     {1-\rho_s\,e^{-i\Phi_s(\omega_a)}},
\label{eq:signal_circ_clean}
\end{equation}
where $\Phi_s(\omega_a)$ is the round-trip phase of the $s$-polarized signal
relative to the carrier resonance. Writing the round-trip time as
$\tau\equiv 2L/c$, we parameterize this phase as
\begin{equation}
\Phi_s(\omega_a)=\omega_a\tau+\Phi_{\rm pol},
\label{eq:Phi_s_def_clean}
\end{equation}
where $\Phi_{\rm pol}$ denotes the polarization-dependent contribution from the
round-trip polarization map (for either the PSM or QWP implementation).

The signal field transmitted through the output coupler is
\begin{equation}
E_s^{(\mathrm{out})}(\omega_a)=t_2^s\,E_s^{(\mathrm{circ})}(\omega_a),
\label{eq:signal_out_clean}
\end{equation}
with $t_{2s}$ the output-coupler amplitude transmission for $s$ polarization.
Combining Eqs.~\eqref{eq:pump_circ_clean}--\eqref{eq:signal_out_clean} gives
\begin{equation}
\frac{E_s^{(\mathrm{out})}(\omega_a)}{E_{0,p}}
=
\beta_0(\omega_a,L)\,
\frac{t_{1}^{p}\,t_{2}^{s}}
{(1-\rho_p)\,\bigl[1-\rho_s e^{-i\Phi_s(\omega_a)}\bigr]} .
\label{eq:transfer_general_clean}
\end{equation}

The transmitted carrier field is
\begin{equation}
E_p^{(\mathrm{out})}
=
t_2^{p}E_p^{(\mathrm{circ})}
=
\frac{t_1^{p}t_2^{p}}{1-\rho_p}\,E_{0,p}.
\label{eq:transmitted_carrier_clean}
\end{equation}
Dividing Eq.~\eqref{eq:transfer_general_clean} by
Eq.~\eqref{eq:transmitted_carrier_clean} gives the orthogonal field ratio at the
cavity output,
\begin{equation}
\frac{E_s^{(\mathrm{out})}(\omega_a)}
{E_p^{(\mathrm{out})}}
=
\beta_0(\omega_a,L)\,
\frac{t_2^s}{t_2^p}\,
\frac{1}{1-\rho_s e^{-i\Phi_s(\omega_a)}} .
\label{eq:orthogonal_output_ratio_clean}
\end{equation}
The PEM readout measures the quadrature component of this orthogonal field. We
therefore define the rotation-to-ellipticity transfer coefficient by
\begin{equation}
\Psi(\omega_a)
=
\kappa_{\beta\to\Psi}(\omega_a)\,\beta(\omega_a),
\label{eq:kappa_beta_to_psi_def}
\end{equation}
with
\begin{equation}
\kappa_{\beta\to\Psi}(\omega_a)
=
\left|
\frac{t_2^s}{t_2^p}
\right|
\frac{1}
{\left|1-\rho_s e^{-i\Phi_s(\omega_a)}\right|}.
\label{eq:kappa_beta_to_psi_response}
\end{equation}
Here the demodulation phase is chosen to read out the signal quadrature. For
\(t_2^s\simeq t_2^p\), this coefficient is determined by the signal-mode cavity
response alone.

\subsection{High-finesse approximation and impedance matching}

For \(T_i\ll 1\) and small round-trip \emph{power} loss
\(l_{\mathrm{int,rt},j}\) in polarization \(j\in\{s,p\}\), one has
\begin{equation}
1-\rho_j \simeq
\frac{T_1+T_2+l_{\mathrm{int,rt},j}}{2}.
\label{eq:1minusR_approx_clean}
\end{equation}
The on-resonance circulating \emph{power} buildup is therefore
\begin{equation}
B_j \equiv |G_j|^2
\simeq
\frac{4T_1}
{\bigl(T_1+T_2+l_{\mathrm{int,rt},j}\bigr)^2}.
\label{eq:power_buildup}
\end{equation}

For the signal channel,
\begin{align}
\left|1-\rho_s e^{-i\Phi_s}\right|^2
&=
(1-\rho_s)^2
+
4\rho_s\sin^2\!\left(\frac{\Phi_s}{2}\right)
\nonumber\\
&\simeq
\left(
\frac{T_1+T_2+l_{\mathrm{int,rt},s}}{2}
\right)^2
+
4\sin^2\!\left(\frac{\Phi_s}{2}\right),
\label{eq:denom_detuned_clean}
\end{align}
where the final expression uses \(\rho_s\simeq1\). This makes explicit that the
signal response is suppressed when the axion sideband is detuned from the
\(s\)-mode resonance.

For each cavity configuration, maximizing the extracted signal typically favors
approximate impedance matching in the \(s\) channel,
\begin{equation}
T_1 \simeq T_2+l_{\mathrm{int,rt},s}.
\label{eq:impedance_match_clean}
\end{equation}
In Table~\ref{tab:cavity_specs}, \(T_1\) and \(T_2\) are power transmissivities,
and \(l_{\mathrm{int,rt},j}\) is the total round-trip internal power loss used
for polarization \(j\in\{s,p\}\). The quantities \(l_{\mathrm{int,rt},j}\)
exclude the input- and output-coupler transmissions and include the other
round-trip losses included in the cavity model. The input-coupler transmission is
therefore different for the QWP and PS-mirror configurations. 

For the QWP comparison, the available vendor data specify residual AR-coating
reflectance for the two wave-plate surfaces rather than a polarization-dependent
loss. We therefore use the same effective round-trip loss for the two
polarizations. For one wave-plate traversal,
\[
l_{\rm QWP,pass}\simeq25~\mathrm{ppm}+346~\mathrm{ppm}
=371~\mathrm{ppm}.
\]
With two QWPs, each traversed twice per round trip, this gives
\[
l_{\mathrm{int,rt},s}^{\rm QWP}
=
l_{\mathrm{int,rt},p}^{\rm QWP}
\simeq
4l_{\rm QWP,pass}
=
1484~\mathrm{ppm}.
\]

For the PS-mirror cavity, the measured single-reflection transmissions are
\(T_s^{\rm PSM}\simeq0.1~\mathrm{ppm}\) and
\(T_p^{\rm PSM}\simeq19~\mathrm{ppm}\). Since the folded cavity contains four
PSM reflections per round trip, the transmission-only contributions are
\(0.4~\mathrm{ppm}\) and \(76~\mathrm{ppm}\), respectively. Dedicated scatter
and absorption measurements of the fabricated coated mirrors have not yet been
performed. For the design-level loss budget, we therefore add a conservative
non-transmission loss of \(2~\mathrm{ppm}\) per PSM reflection. The total
round-trip internal losses used in Table~\ref{tab:cavity_specs} are therefore
\[
\begin{aligned}
l_{\mathrm{int,rt},s}^{\rm PSM}
&\simeq 4(0.1+2)~\mathrm{ppm}
= 8.4~\mathrm{ppm},\\
l_{\mathrm{int,rt},p}^{\rm PSM}
&\simeq 4(19+2)~\mathrm{ppm}
= 84~\mathrm{ppm}.
\end{aligned}
\]

The numerical parameters used for the sensitivity projections are summarized in
Table~\ref{tab:cavity_specs}. For the dark-matter normalization, we adopt the
standard local density benchmark
\(\rho_{\rm DM}=0.4~\mathrm{GeV}/\mathrm{cm}^{3}\)
\cite{Catena:2009mf,deSalas:2020hbh}.

\begin{table}[t]
\squeezetable
\caption{Main parameters used in the sensitivity analysis. Buildup denotes the
on-resonance \emph{circulating power} enhancement for each polarization and is
dimensionless. The quoted internal losses are total round-trip power losses,
excluding the input- and output-coupler transmissions.}
\label{tab:cavity_specs}
\begin{ruledtabular}
\begin{tabular}{lc}
Parameter & Value \\

\multicolumn{2}{c}{\textbf{General and readout parameters}} \\

Input power & \(200~\mathrm{mW}\) \\
Cavity length & \(1.5~\mathrm{m}\) \\
Analyzer extinction ratio, \(\sigma^2\) & \(10^{-7}\) \\
PEM modulation depth, \(\eta_0\) & \(10^{-2}\) \\
Photodiode quantum efficiency, \(\eta_{\rm QE}\) & \(0.8\) \\
Local dark-matter density, \(\rho_{\rm DM}\) &
\(0.4~\mathrm{GeV}/\mathrm{cm}^{3}\) \\
Coherent integration time & \(t_{\rm coh}=10^6/\nu_a\) \\

\multicolumn{2}{c}{\textbf{QWP cavity}} \\
\(T_1^{\rm QWP}\) & \(1544~\mathrm{ppm}\) \\
\(T_2^{\rm QWP}\) & \(60~\mathrm{ppm}\) \\
\(l_{\rm int,rt,s}^{\rm QWP}\) & \(1484~\mathrm{ppm}\) \\
\(l_{\rm int,rt,p}^{\rm QWP}\) & \(1484~\mathrm{ppm}\) \\
Buildup (\(s\)) & \(648\) \\
Buildup (\(p\)) & \(648\) \\

\multicolumn{2}{c}{\textbf{PS-mirror cavity}} \\
\(T_1^{\rm PSM}\) & \(68.4~\mathrm{ppm}\) \\
\(T_2^{\rm PSM}\) & \(60~\mathrm{ppm}\) \\
\(l_{\rm int,rt,s}^{\rm PSM}\) & \(8.4~\mathrm{ppm}\) \\
\(l_{\rm int,rt,p}^{\rm PSM}\) & \(84~\mathrm{ppm}\) \\
Buildup (\(s\)) & \(14600\) \\
Buildup (\(p\)) & \(6060\) \\
\end{tabular}
\end{ruledtabular}
\end{table}
\subsection{Photon-counting benchmark}

A time-dependent axion background produces circular birefringence: the propagation
eigenmodes are right- and left-circular polarizations,
\begin{equation}
\hat{\mathbf e}_\pm
=
\frac{\hat{\mathbf x}\pm i\hat{\mathbf y}}{\sqrt2},
\end{equation}
which acquire different phases $\phi_\pm$ over one pass. A linearly polarized
input field,
\begin{equation}
\mathbf E_{\rm in}=E_0\,\hat{\mathbf x}
=
\frac{E_0}{\sqrt2}\bigl(\hat{\mathbf e}_+ + \hat{\mathbf e}_-\bigr),
\end{equation}
therefore becomes
\begin{equation}
\mathbf E_{\rm out}
=
E_0 e^{i\bar\phi}
\Bigl[\hat{\mathbf x}\cos\beta+\hat{\mathbf y}\sin\beta\Bigr],
\label{eq:Eout_rotation_clean}
\end{equation}
with
\begin{equation}
\bar\phi\equiv\frac{\phi_++\phi_-}{2},
\qquad
\beta\equiv\frac{\Delta\phi}{2},
\qquad
\Delta\phi\equiv\phi_+-\phi_-.
\label{eq:beta_def_deltaphi_clean}
\end{equation}
For $|\beta|\ll1$, the orthogonal field is $E_\perp\simeq \beta E_0$.

To state the fundamental photon-counting limit in a readout-independent way, we
consider an ideal Stokes-parameter measurement at mid-fringe, so that
\begin{equation}
\langle\Delta N\rangle
\equiv
\langle N_1-N_2\rangle
=
N_{\rm det}\sin(2\beta)
\simeq
2\beta N_{\rm det},
\label{eq:DeltaN_linear_beta_clean}
\end{equation}
where $N_{\rm det}\equiv \langle N_1+N_2\rangle$ is the total detected photon
number during an integration time $t_{\rm int}$. For coherent light,
$\mathrm{Var}(\Delta N)\simeq N_{\rm det}$, so the shot-noise-limited minimum
detectable rotation for ${\rm SNR}=1$ is
\begin{equation}
\beta_{\min}(t_{\rm int})=\frac{1}{2\sqrt{N_{\rm det}}}.
\label{eq:beta_min_SQL_counts_clean}
\end{equation}
With detected photon flux
\begin{equation}
\dot N_{\rm det}=\frac{\eta_{\rm QE} P_{\rm det}}{\hbar\omega_0},
\end{equation}
this becomes the single-sided amplitude spectral density
\begin{equation}
N_{\beta}^{\rm shot}
=
\frac{1}{2\sqrt{\dot N_{\rm det}}}
=
\sqrt{\frac{\hbar\omega_0}{4\,\eta_{\rm QE}\,P_{\rm det}}}
\;\;{\rm rad}/\sqrt{\rm Hz},
\label{eq:Nbeta_shot_power_clean}
\end{equation}
where \(\eta_{\rm QE}\) is the photodiode quantum efficiency and
\(P_{\rm det}\) is the optical power incident on the detector for the chosen
estimator. Any downstream optical losses may be folded into \(P_{\rm det}\).
The photon energy is included through
\(\hbar\omega_0=hc/\lambda_0\). Equation~\eqref{eq:Nbeta_shot_power_clean}
therefore gives a readout-independent photon-counting benchmark for an ideal
Stokes-parameter measurement. It does not include the specific PEM modulation
scheme or technical noise terms, which are treated separately in
Sec.~\ref{sec:heterodyne_detection}.

Assuming that axions saturate the local dark-matter density, the field amplitude
$a_0$ is fixed by
\begin{equation}
\rho_{\rm DM}=\frac12 m_a^2 a_0^2
\qquad
\text{(natural units in the axion sector)}.
\label{eq:a0_density_relation}
\end{equation}
Equating the cavity-enhanced output polarization signal to the shot-noise
benchmark then gives
\begin{equation}
g_{a\gamma\gamma}^{\min}(\omega_a)
=
\frac{1}{\sqrt{t_{\rm int}}}\,
\frac{N_{\beta}^{\rm shot}}
     {a_0\,\bigl|\sin(\omega_a L/2)\bigr|\,
      \bigl|\kappa_{\beta\to\Psi}(\omega_a)\bigr|},
\label{eq:sensitivity_general_clean}
\end{equation}
where \(\kappa_{\beta\to\Psi}\), defined in
Eq.~\eqref{eq:kappa_beta_to_psi_response}, is the cavity rotation-to-ellipticity transfer
coefficient. It converts the one-pass axion-induced rotation into the
polarization signal measured at the cavity output. This is the appropriate
normalization for the photon-counting benchmark because the ideal Stokes
measurement is sensitive to the orthogonal field relative to the transmitted
carrier.

\begin{figure}[t]
    \centering
    \includegraphics[width=0.95\linewidth]{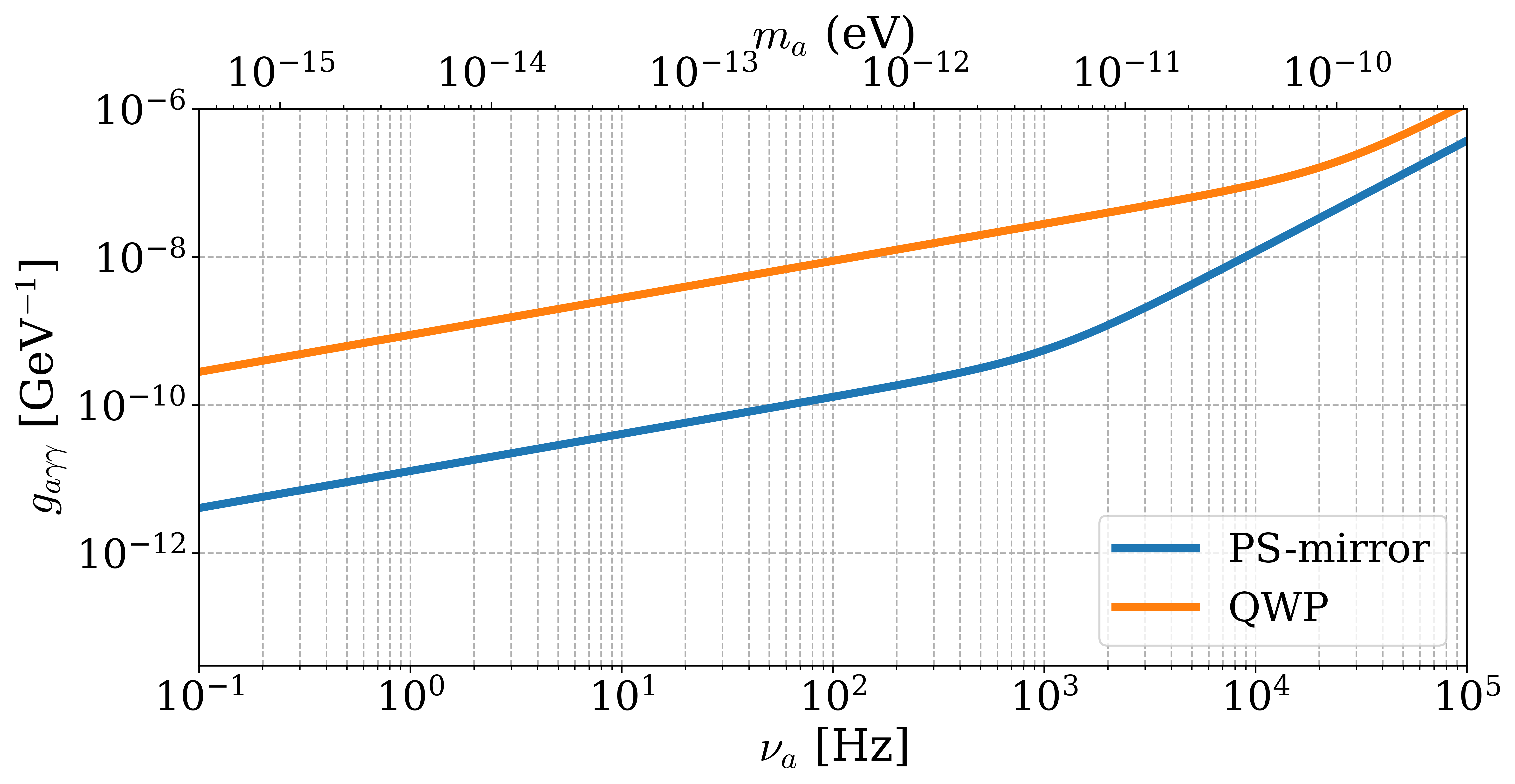}
    \caption{\small
Projected photon-counting sensitivity to the axion--photon coupling
\(g_{a\gamma\gamma}\) versus axion Compton frequency \(\nu_a\) and mass for the
two cavity configurations studied here. The orange curve shows the cavity with
two intracavity quarter-wave plates (QWPs), while the blue curve shows the
phase-shifting-mirror cavity. The curves assume the cavity parameters of
Table~\ref{tab:cavity_specs}, input power \(P_0=200~\mathrm{mW}\), local dark
matter density \(\rho_{\rm DM}=0.4~\mathrm{GeV}/\mathrm{cm}^{3}\), and coherent
integration time \(t_{\rm int}=t_{\rm coh}=10^6/\nu_a\). Only the ideal
photon-counting limit is included in this benchmark; coating-birefringence noise,
PEM technical readout noise, and angular-jitter coupling are not included.
}
    \label{fig:SENSITIVITY}
\end{figure}

In the short-baseline limit \(\omega_aL\ll1\),
\begin{equation}
g_{a\gamma\gamma}^{\min}(\omega_a)
\simeq
\frac{1}{\sqrt{t_{\rm int}}}\,
\frac{2N_{\beta}^{\rm shot}}
     {a_0\,\omega_a L\,
      \bigl|\kappa_{\beta\to\Psi}(\omega_a)\bigr|}.
\label{eq:sensitivity_small_omegaL_clean}
\end{equation}
These expressions assume coherent integration at known frequency and phase. For
the projections shown here, we take the coherent integration time to be
\(t_{\rm int}=t_{\rm coh}=10^6/\nu_a\), as summarized in
Table~\ref{tab:cavity_specs}. Using
Eqs.~\eqref{eq:sensitivity_general_clean} and
\eqref{eq:sensitivity_small_omegaL_clean} together with the cavity parameters of
Table~\ref{tab:cavity_specs}, we obtain the projected sensitivities shown in
Fig.~\ref{fig:SENSITIVITY}.

As expected from the lower round-trip loss budget of the reflective implementation,
the phase-shifting-mirror cavity provides a markedly improved low-mass reach relative to
the QWP configuration.

\subsection{Frequency detuning via angle-dependent phase shifts}
\label{sec:detuning}

The phase-shifting mirrors exhibit an angle-dependent differential reflection phase
\(
\Delta\phi(\theta)=\phi_s(\theta)-\phi_p(\theta)
\)
(Fig.~\ref{fig:phase_simulation}). In the folded cavity, this coating phase contributes to
the round-trip phase of the $s$-polarized signal mode. We therefore write the
$s$-mode round-trip phase at axion sideband offset frequency $\omega_a$ as
\begin{equation}
\Phi_s(\omega_a,\theta)=\omega_a\tau+\Phi_{\mathrm{PSM}}(\theta),
\label{eq:Phi_s_theta}
\end{equation}
where $\tau\equiv 2L/c$ is the cavity round-trip time and
$\Phi_{\mathrm{PSM}}(\theta)$ denotes the polarization-dependent contribution from the
PS-mirror round-trip polarization map.

For small AOI excursions about the operating point $\theta_0$, one may linearize
\begin{equation}
\Phi_{\mathrm{PSM}}(\theta)\simeq
\Phi_{\mathrm{PSM}}(\theta_0)
+
\left.\frac{\partial\Phi_{\mathrm{PSM}}}{\partial\theta}\right|_{\theta_0}\delta\theta,
\qquad
\delta\theta\equiv \theta-\theta_0.
\label{eq:dPhi_dtheta}
\end{equation}
If the two PSMs are tuned independently, the corresponding generalization is
\begin{equation}
\delta\Phi_{\mathrm{PSM}}=
\frac{\partial\Phi_{\mathrm{PSM}}}{\partial\theta_3}\,\delta\theta_3
+
\frac{\partial\Phi_{\mathrm{PSM}}}{\partial\theta_4}\,\delta\theta_4.
\label{eq:dPhi_dtheta_two}
\end{equation}

The resulting shift of the $s$-mode resonance frequency follows from the resonance
condition \(\Phi_s(\omega)=2\pi m\):
\begin{equation}
\delta\omega_s\,\tau+\delta\Phi_{\mathrm{PSM}}=0,
\qquad
\delta\omega_s=-\frac{\delta\Phi_{\mathrm{PSM}}}{\tau},
\label{eq:detuning_dw}
\end{equation}
or, in ordinary frequency,
\begin{equation}
\delta\nu_s
=
-\frac{\mathrm{FSR}}{2\pi}\,\delta\Phi_{\mathrm{PSM}},
\qquad
\mathrm{FSR}\equiv \frac{1}{\tau}.
\label{eq:detuning_dnu}
\end{equation}
Thus AOI tuning provides a direct handle on the signal-mode detuning.

In practice, detuning can be implemented by varying the AOI of one PSM while keeping
the other fixed. Coordinated tuning of both mirrors can reduce beam walk-off and
alignment perturbations. An analogous mechanism exists in the QWP cavity, where the
relative orientation of the two QWPs modifies the round-trip polarization map and can
split the polarization eigenfrequencies. However, QWP-axis tuning also changes the
cavity eigenpolarizations and introduces mode mixing, so we do not use it as the
primary tuning method in the present sensitivity estimates.

The cavity is locked to the \(p\)-polarized pump by the
Pound--Drever--Hall technique~\cite{Drever:1983qsr,Black:2001jtc}.
The feedback loop suppresses common-mode detuning by keeping the pump on resonance, but
it does not cancel the AOI-induced \emph{differential} phase shift of the $s$ mode.
Therefore, to leading order, AOI tuning shifts the signal resonance relative to the
locked pump while leaving the pump resonance condition unchanged.

\begin{figure}[!tbp]
\centering
\includegraphics[width=0.95\linewidth]{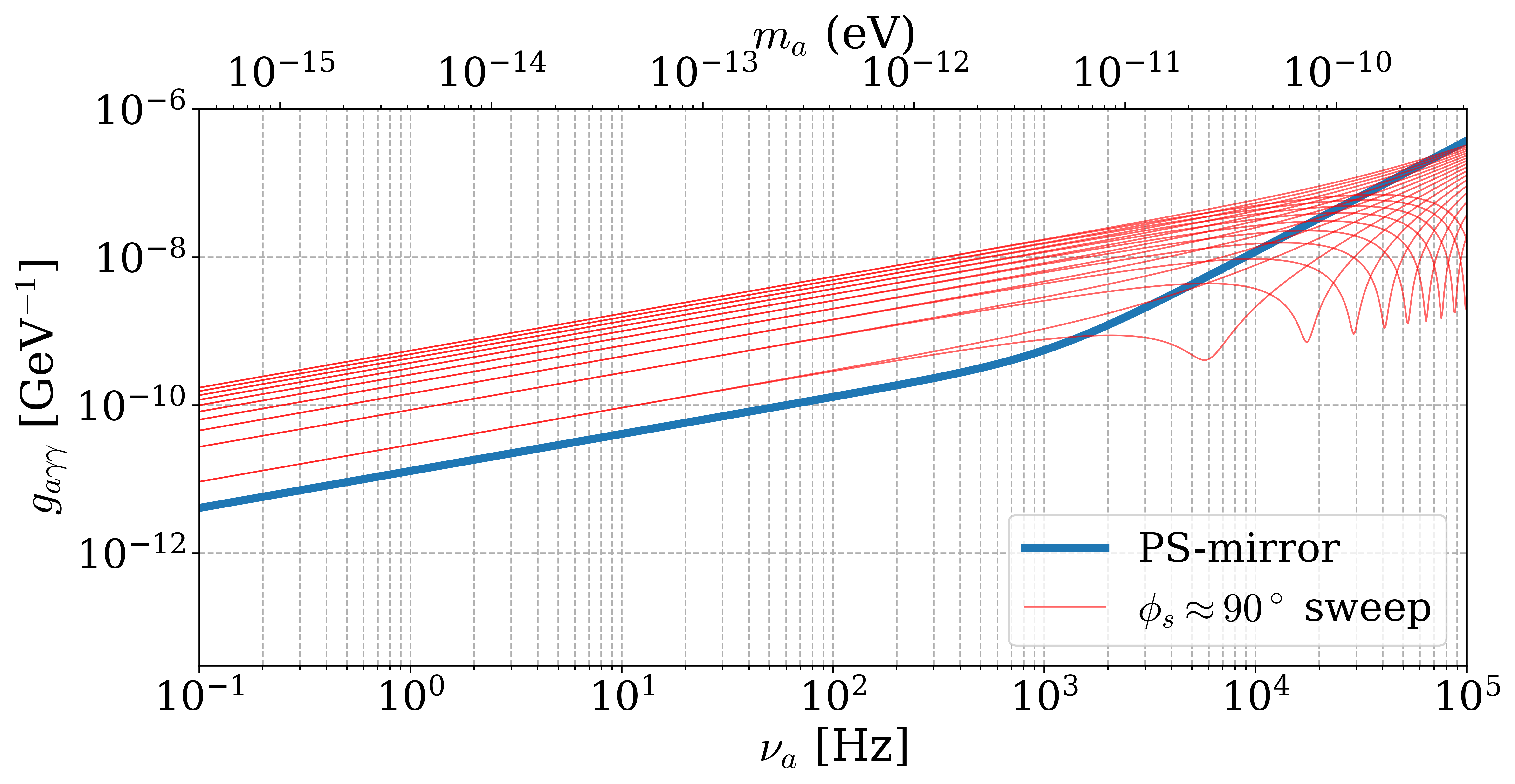}
\caption{\small
Projected photon-counting sensitivity to the axion--photon coupling
\(g_{a\gamma\gamma}\) as a function of axion Compton frequency \(\nu_a\) and
mass for different detuning conditions of the signal-polarization cavity mode.
The blue curve shows the reference case in which the signal mode is on
resonance. The family of red curves shows the sensitivity obtained when the
angle of incidence of the phase-shifting mirrors is varied, thereby changing the
birefringent round-trip phase and shifting the \(s\)-polarized cavity resonance
relative to the locked pump. Each trace corresponds to a distinct detuning
setting, which moves the region of maximum sensitivity across the frequency
band. The curves assume the cavity parameters of Table~\ref{tab:cavity_specs},
input power \(P_0=200~\mathrm{mW}\), local dark-matter density
\(\rho_{\rm DM}=0.4~\mathrm{GeV}/\mathrm{cm}^{3}\), and coherent integration
time \(t_{\rm int}=t_{\rm coh}=10^6/\nu_a\). Only the ideal photon-counting
limit is included; PEM technical readout noise, coating-birefringence noise, and
angular-jitter coupling are not included.}
\label{fig:axion_sensitivity_detuning}
\end{figure}

This tuning enters the sensitivity model through the signal-mode denominator
\begin{equation}
\left|1-\rho_s e^{-i\Phi_s(\omega_a,\theta)}\right|,
\label{eq:signal_denominator_detuned}
\end{equation}
or equivalently through the angle-dependent rotation-to-ellipticity response
\begin{equation}
\kappa_{\beta\to\Psi}(\omega_a,\theta)
=
\left|
\frac{t_2^s}{t_2^p}
\right|
\frac{1}
{\left|1-\rho_s e^{-i\Phi_s(\omega_a,\theta)}\right|}.
\label{eq:kappa_beta_to_psi_detuned}
\end{equation}
Over the small tuning range of interest, we neglect the residual angle
dependence of the amplitude factors and retain only the phase dependence in
\(\Phi_s(\omega_a,\theta)\). The resulting redistribution of sensitivity across
the search band is illustrated in Fig.~\ref{fig:axion_sensitivity_detuning}.
AOI tuning does not improve the sensitivity uniformly, but rather shifts the
region of maximum response to different axion frequencies. The conversion of the
transmitted orthogonal field into a measured ellipticity, and its demodulation
with the PEM, are described in Sec.~\ref{sec:heterodyne_detection}.

\section{Heterodyne Detection Technique}
\label{sec:heterodyne_detection}

Section~\ref{section-sensitivity} established the readout-independent relation between the
axion-induced cavity signal and the fundamental photon-counting limit. Here we specify the
PEM-based implementation used to read out the small polarization signal at the cavity output and
quantify the additional technical noises associated with that implementation.

The beam emerging from the cavity carries a small polarization signal at the axion frequency
\(\nu_a=\omega_a/2\pi\). Direct measurement at \(\nu_a\) is generally unfavorable because this
frequency can lie in a region of elevated technical noise. We therefore up-convert the signal
with a photoelastic modulator (PEM), driven at a fixed frequency
\(\nu_{\rm PEM}\simeq 50~\mathrm{kHz}\), and recover the signal by lock-in demodulation around
\(\nu_{\rm PEM}\).

The PEM imposes a known time-dependent ellipticity
\begin{equation}
\eta(t)=\eta_0\cos(2\pi\nu_{\rm PEM} t),
\label{eq:PEM_signal}
\end{equation}
with modulation depth \(\eta_0\ll1\). Let \(\Psi(t)\) denote the physical ellipticity of the
beam incident on the analyzer. We write the analyzer-input ellipticity in terms of the axion-induced rotation as
\begin{equation}
\Psi(\omega_a)
=
\kappa_{\beta\to\Psi}(\omega_a)\,\beta(\omega_a),
\label{eq:kappa_beta_to_psi}
\end{equation}
where \(\kappa_{\beta\to\Psi}\) is the cavity-enhanced rotation-to-ellipticity
transfer coefficient defined in Eq.~\eqref{eq:kappa_beta_to_psi_response}. This coefficient is
normalized to the transmitted carrier field and includes the signal-mode buildup
and the readout quadrature selected by the demodulation phase.

At the field level, the PEM-generated ellipticity and the physical cavity ellipticity enter as a
quadrature component of the transmitted beam. Near extinction, the detected power therefore
depends on the modulus squared of the total small ellipticity:
\begin{equation}
\begin{aligned}
\frac{P_{\rm det}(t)}{P_{\rm carr}}
&\simeq
\sigma^2+\bigl(\eta(t)+\Psi(t)\bigr)^2 \\
&\simeq
\sigma^2+\eta^2(t)+2\eta(t)\Psi(t),
\end{aligned}
\label{eq:darkport_power_clean}
\end{equation}
where \(P_{\rm carr}\) is the transmitted carrier power incident on the analyzer,
\(\sigma^2\) is the residual extinction ratio, and terms of order \(\Psi^2\)
have been neglected. If \(P_0\) denotes the input laser power incident on the
cavity, then, up to downstream optical losses,
\begin{equation}
P_{\rm carr}
=
P_0
\left|
\frac{t_1^p t_2^p}{1-\rho_p}
\right|^2 .
\label{eq:Pin_analyzer}
\end{equation}
The signal-mode enhancement is not included in \(P_{\rm carr}\); it enters the
heterodyne readout through \(\kappa_{\beta\to\Psi}\).

Using Eq.~\eqref{eq:PEM_signal}, the quadratic PEM term can be written as
\[
\eta^2(t)=\frac{\eta_0^2}{2}\Bigl[1+\cos(4\pi\nu_{\rm PEM}t)\Bigr].
\]
Accordingly, the detected power contains a DC component
\begin{equation}
P_{\rm dc}=P_{\rm carr}\left(\sigma^2+\frac{\eta_0^2}{2}\right),
\label{eq:Pdc_def}
\end{equation}

and a second-harmonic component at \(2\nu_{\rm PEM}\) with amplitude
\[
P_{2\nu_{\rm PEM}}=\frac{P_{\rm carr}\eta_0^2}{2}.
\]
The latter is retained because it provides an in situ calibration of the PEM modulation depth
\(\eta_0\), while the cross term \(2\eta(t)\Psi(t)\) mixes the physical ellipticity to sidebands
around \(\nu_{\rm PEM}\). Demodulation at \(\nu_{\rm PEM}\) therefore yields a calibrated
estimate of \(\Psi(t)\), and hence of \(\beta(t)\).

\begin{figure}[t]
  \centering
  \includegraphics[width=1\linewidth]{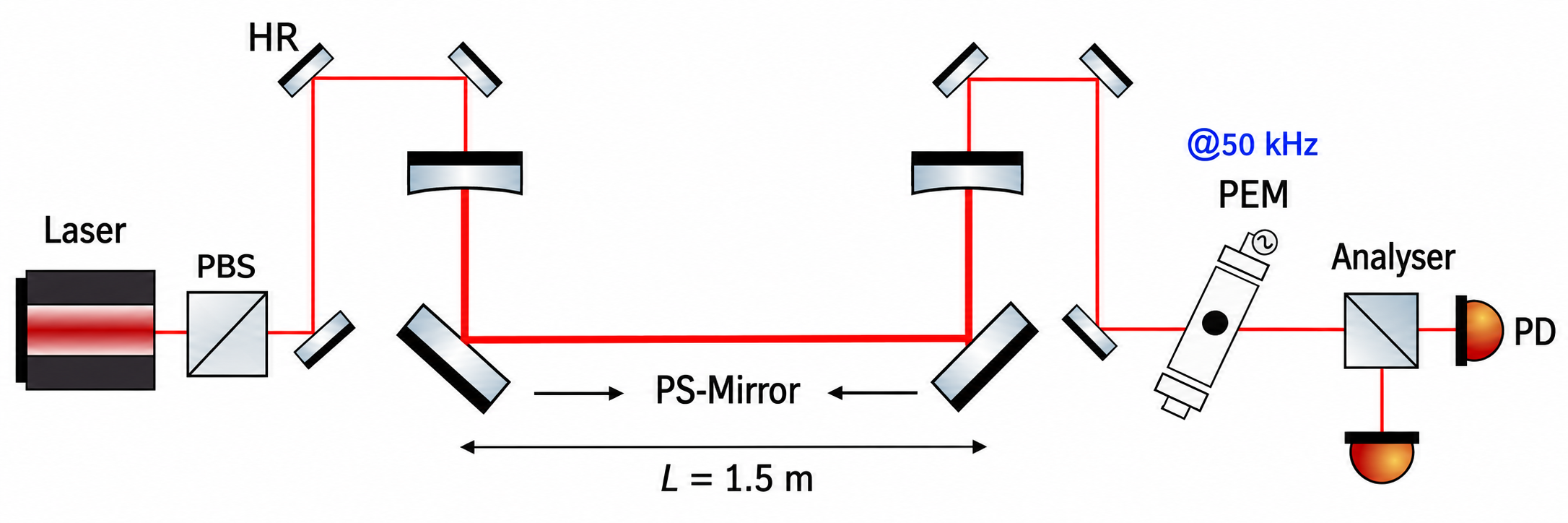}
 \caption{Schematic of the APE cavity polarimeter with PEM-based heterodyne readout. A
linearly polarized laser is injected through a polarizer into the folded
cavity containing two phase-shifting mirrors. The beam transmitted by the cavity passes through
the photoelastic modulator (PEM), which imposes a known ellipticity modulation at
\(\nu_{\rm PEM}\), and is then analyzed and detected on a photodiode (PD). The PEM therefore
provides the local oscillator that mixes the small cavity-output polarization signal into a
narrow band around \(\nu_{\rm PEM}\) for lock-in detection.}
  \label{fig:fullsetup}
\end{figure}

\subsection{Readout-noise model and modulation-depth dependence}
\label{noise}
At the detector, the heterodyne signal arises from the cross term
\(2\eta(t)\Psi(t)\). For
\(\eta(t)=\eta_0\cos(2\pi\nu_{\rm PEM}t)\) and
\(\Psi(t)=\Psi_0\cos(2\pi\nu_a t)\), this term produces sidebands at
\(\nu_{\rm PEM}\pm\nu_a\),
\begin{equation}
P_{\rm sig}(t)
=
P_{\rm carr}\eta_0\Psi_0
\Bigl[
\cos 2\pi(\nu_{\rm PEM}+\nu_a)t
+
\cos 2\pi(\nu_{\rm PEM}-\nu_a)t
\Bigr].
\label{eq:heterodyne_sidebands}
\end{equation}
Thus the amplitude of each first-order sideband is
\(P_{\rm carr}\eta_0\Psi_0\). Equivalently, after lock-in demodulation at
\(\nu_{\rm PEM}\), the baseband signal is linear in \(\Psi\) with slope
\begin{equation}
\frac{\partial P_{\rm sig}}{\partial \Psi}\simeq P_{\rm carr}\eta_0.
\label{eq:dPdPsi}
\end{equation}
Using \(\Psi(\omega_a)=\kappa_{\beta\to\Psi}(\omega_a)\beta(\omega_a)\), the
power-to-rotation conversion is therefore
\begin{equation}
N_\beta=
\frac{N_P}{P_{\rm carr}\eta_0\,|\kappa_{\beta\to\Psi}(\omega_a)|},
\label{eq:Nbeta_from_NP_clean}
\end{equation}
where \(N_P\) is the single-sided power-noise ASD referred to the heterodyne signal band.

It is convenient to write the photodiode responsivity as \(R_{\rm PD}\) (A/W),
while keeping the quantum efficiency explicit as \(\eta_{\rm QE}\). The optical power \(P_{\rm carr}\) is the transmitted carrier power defined in
Eq.~\eqref{eq:Pin_analyzer}, and
\(\kappa_{\beta\to\Psi}\) is the rotation-to-ellipticity transfer coefficient
defined in Eq.~\eqref{eq:kappa_beta_to_psi_response}.

\paragraph*{Shot noise.}
The photocurrent is \(i_{\rm dc}=R_{\rm PD}P_{\rm dc}\), so the corresponding
single-sided shot-noise power ASD is
\begin{equation}
N_P^{\rm shot}
=
\frac{\sqrt{2e\,i_{\rm dc}}}{R_{\rm PD}}
=
\sqrt{\frac{2\hbar\omega_0 P_{\rm dc}}{\eta_{\rm QE}}}.
\label{eq:NP_shot_clean}
\end{equation}
The equivalent rotation-noise ASD is therefore
\begin{equation}
N_\beta^{\rm shot}
=
\frac{1}{\eta_0\,|\kappa_{\beta\to\Psi}(\omega_a)|}
\sqrt{\frac{2\,\hbar\omega_0}{\eta_{\rm QE}\,P_{\rm carr}}
\left(\sigma^2+\frac{\eta_0^2}{2}\right)}.
\label{eq:Nbeta_shot_clean}
\end{equation}
In the LO-dominated regime, \(\eta_0\gg\sigma\), the explicit \(\eta_0\) dependence
cancels and the shot-noise floor approaches a constant.

\paragraph*{Electronics noise.}
Let \(i_n^{\rm elec}\) denote the total input-referred current-noise ASD of the
detection chain at \(\nu_{\rm PEM}\). If Johnson noise of the transimpedance
resistor dominates, then \(i_n^{\rm elec}\simeq \sqrt{4k_B T/R_f}\), but the same
notation also includes amplifier current or voltage noise. The equivalent
rotation-noise ASD is
\begin{equation}
N_\beta^{\rm elec}
=
\frac{i_n^{\rm elec}}
{R_{\rm PD}P_{\rm carr}\eta_0\,|\kappa_{\beta\to\Psi}(\omega_a)|}.
\label{eq:Nbeta_elec_clean}
\end{equation}

\paragraph*{Dark-current noise.}
If the photodiode dark current is \(i_{\rm dark}\), its shot-noise contribution gives
\begin{equation}
N_\beta^{\rm dark}
=
\frac{\sqrt{2e\,i_{\rm dark}}}
{R_{\rm PD}P_{\rm carr}\eta_0\,|\kappa_{\beta\to\Psi}(\omega_a)|}.
\label{eq:Nbeta_dark_clean}
\end{equation}

\paragraph*{Relative intensity noise.}
Laser intensity fluctuations near the PEM frequency generate power noise at the
detector. Because the PEM-induced term \(\eta^2(t)\) contains both a DC
component and a component at \(2\nu_{\rm PEM}\), both contributions can mix
intensity noise into the heterodyne band. Treating the two contributions as
statistically independent, the resulting power-noise ASD is
\begin{equation}
N^{\rm RIN}_{P}
=
{\rm RIN}(\nu_{\rm PEM})\,P_{\rm carr}
\left[
\left(\sigma^2+\frac{\eta_0^2}{2}\right)^2
+
\left(\frac{\eta_0^2}{2}\right)^2
\right]^{1/2}.
\label{eq:rin_power_noise}
\end{equation}
This corresponds to
\begin{equation}
N^{\rm RIN}_{\beta}
=
\frac{{\rm RIN}(\nu_{\rm PEM})}
{\eta_0 |\kappa_{\beta\to\Psi}(\omega_a)|}
\left[
\left(\sigma^2+\frac{\eta_0^2}{2}\right)^2
+
\left(\frac{\eta_0^2}{2}\right)^2
\right]^{1/2}.
\label{eq:rin_rotation_noise}
\end{equation}

Assuming these contributions are statistically independent, the total readout-noise
ASD is
\begin{equation}
N_\beta^{\rm tot}
=
\left[
\bigl(N_\beta^{\rm shot}\bigr)^2
+\bigl(N_\beta^{\rm elec}\bigr)^2
+\bigl(N_\beta^{\rm dark}\bigr)^2
+\bigl(N_\beta^{\rm RIN}\bigr)^2
\right]^{1/2}.
\label{eq:noise_budget_beta}
\end{equation}

The dependence on the PEM modulation depth \(\eta_0\) is then straightforward: at
small \(\eta_0\), shot noise, electronics noise, and dark-current noise scale
approximately as \(1/\eta_0\), whereas the RIN contribution increases once the local
oscillator dominates the DC detector power. In the present design, we adopt
\(\eta_0=(1-10)~\mathrm{mrad}\), for which the total readout noise is close to the
shot-noise floor. We therefore use \(N_\beta^{\rm tot}\) evaluated at this operating
point in the sensitivity projections and do not show a separate modulation-depth
optimization figure.

These readout noises do not exhaust the full sensitivity budget. Coating-birefringence fluctuations and angular motion of the optics perturb the
physical cavity polarization state before the PEM and therefore are not
suppressed by increasing the artificial local-oscillator amplitude. We discuss those low-frequency limitations in
Sec.~\ref{subsec:open_biref_discussion}.

\subsection{Intrinsic coating-birefringence benchmark}
\label{subsec:open_biref_discussion}

At present there is no complete microscopic model that predicts coating-induced
polarization noise for the APE cavity at the level required for a closed noise
budget. We therefore use the residual polarimetric noise measured in a
high-finesse precision polarimeter as an empirical reference spectrum
\cite{ejlli2020pvlas}. This choice is useful because such systems directly
measure residual ellipticity noise relevant to cavity polarimetry. However, the
extrapolation to APE is not a first-principles prediction: the systems differ in
coating design, angle of incidence, polarization eigenmodes, cavity geometry, and
mechanical environment. The model below should therefore be interpreted as an
illustrative coating-birefringence benchmark for the design study, not as a
demonstrated noise floor for the assembled APE instrument.

We work directly with the equivalent optical-path-difference ASD of the
reference polarimeter, denoted by \(N^{\rm ref}_{\Delta D}(\nu)\). We
parameterize this reference OPD spectrum as
\begin{equation}
N^{\rm ref}_{\Delta D}(\nu)
=
\left[
\frac{A_{\rm th}^2}
{\nu\left[1+(\nu/\nu_0)^2\right]}
+
B_{\rm th}^2\nu^{-1/2}
\right]^{1/2},
\label{eq:ref_opd_spectrum}
\end{equation}
with \(A_{\rm th}=(2.01\pm0.02)\times10^{-18}~\mathrm{m}\),
\(B_{\rm th}=(4.63\pm0.02)\times10^{-19}~\mathrm{m\,Hz}^{1/4}\), and
\(\nu_0=(15.0\pm0.4)~\mathrm{Hz}\) \cite{ejlli2020pvlas}. We use this spectrum only as a measured
reference level for residual coating-birefringence noise in a high-finesse
polarimeter.

To translate this reference spectrum to APE, we introduce a dimensionless
coating-birefringence scaling parameter \(\chi_{\rm biref}\),
\begin{equation}
N^{\rm APE}_{\Delta D}(\nu;\chi_{\rm biref})
=
\chi_{\rm biref}\,
N^{\rm ref}_{\Delta D}(\nu).
\label{eq:ape_biref_scaled_opd}
\end{equation}
The nominal choice \(\chi_{\rm biref}=1\) corresponds to using the reference
coating-birefringence level directly for APE. This is a reasonable design
benchmark because the relevant mirror and beam-size scales are comparable, but
it should not be interpreted as a first-principles prediction. We therefore
treat \(\chi_{\rm biref}\) as a phenomenological uncertainty parameter rather
than deriving it from a simple counting of optical surfaces.

The OPD spectrum \(N^{\rm APE}_{\Delta D}\) is then converted to an equivalent
polarization-noise ASD through the phase retardance associated with an optical
path difference.
\begin{equation}
N_{\beta}^{\rm coat}(\nu;\chi_{\rm biref})
\simeq
\frac{\pi}{\lambda}\,
N^{\rm APE}_{\Delta D}(\nu;\chi_{\rm biref}) .
\label{eq:biref_beta_noise}
\end{equation}
Equivalently, the corresponding analyzer-input ellipticity noise is
\begin{equation}
N_{\Psi}^{\rm coat}(\nu;\chi_{\rm biref})
=
\left|\kappa_{\beta\to\Psi}(\nu)\right|\,
N_{\beta}^{\rm coat}(\nu;\chi_{\rm biref}) .
\label{eq:biref_ellipticity_noise}
\end{equation}
Equation~\eqref{eq:biref_ellipticity_noise} is used only when the coating noise
is expressed at the analyzer input. When the noise is referred back to an
equivalent axion-induced rotation, the same factor
\(\kappa_{\beta\to\Psi}\) divides out. Therefore the coating-birefringence
contribution to the coupling sensitivity is obtained from
\(N_{\beta}^{\rm coat}\), not from an additional cavity-enhanced noise term.

In the sensitivity figures, the curve obtained with \(\chi_{\rm biref}=1\)
should be read as the nominal coating-birefringence benchmark. To show the impact
of the uncertain extrapolation, we also consider the range
\(\chi_{\rm biref}=0.5,1,2\). This range is not assigned a statistical confidence
level; it is intended to indicate how strongly the projected low-frequency reach
depends on the assumed coating-birefringence noise. A direct measurement of the
birefringence-noise spectrum of the assembled APE cavity will ultimately be
required before the corresponding sensitivity can be interpreted as a
quantitative detector performance.

\paragraph*{Remark on crystalline coatings.}

Crystalline multilayer coatings, such as GaAs/AlGaAs stacks, can substantially reduce
coating Brownian \emph{phase} noise and improve cavity frequency stability
~\cite{Cole2013TenfoldBrownian}. This does not, however, automatically imply a
corresponding reduction of polarization noise. In particular, Brownian strain can couple
through photoelastic coefficients to fluctuations of the birefringence axes and thereby
generate ellipticity noise~\cite{Zhang2024NABPNoise}. For this reason, alternative
coating technologies should be evaluated by direct polarization-noise measurements, not
only by their phase-noise performance.

\subsection{Angular-jitter coupling and commissioning requirements}
\label{subsec:seismic_coupling}

Angular motion of the cavity optics can couple to the polarimetric readout by
perturbing the round-trip polarization map. In the phase-shifting-mirror
configuration, yaw motion changes the angle of incidence on the folding mirrors
and therefore modulates the differential reflection phase
\(\Delta\phi(\theta)=\phi_s(\theta)-\phi_p(\theta)\). Pitch motion rotates the
local plane of incidence and hence the local \(s/p\) basis of the coating
relative to the laboratory polarization basis. Both effects can convert angular
jitter into an apparent ellipticity at the analyzer input.

A useful way to view this coupling is to distinguish retardance fluctuations
from axis-mixing fluctuations. For a birefringent element with eigenaxes rotated
by an angle \(\alpha\) relative to the incident linear polarization, a small
retardance fluctuation produces an ellipticity proportional to
\(\delta\phi\,\sin(2\alpha)/2\). The coupling is therefore maximal when the
polarization is at \(45^\circ\) to the birefringent axes, but it vanishes to
first order when the polarization is aligned with one eigenaxis. In the ideal
APE operating point, the \(p\)-polarized carrier is chosen to be an
eigenpolarization of the cavity round-trip map. Pure fluctuations of the
differential phase therefore do not directly generate carrier ellipticity at
first order. Residual coupling can nevertheless arise from imperfect
eigenpolarization alignment, pitch-induced rotations of the local \(s/p\) basis,
static misalignments, and higher-order products of angular motion and
birefringence.

This angular-to-polarization coupling is different from the displacement-noise
limitation of gravitational-wave interferometers. Precision cavity polarimeters
are typically discussed in terms of residual ellipticity, intrinsic
birefringence, alignment-to-polarization coupling, and technical polarimetric
noise, rather than a direct seismic displacement background
\cite{ejlli2020pvlas}. For APE, however, the phase-shifting mirrors introduce an
explicit angle-dependent polarization phase, so yaw and pitch motion must be
checked experimentally.

We do not include angular-jitter noise as a quantitative contribution in the
sensitivity projections, because the relevant yaw- and pitch-to-ellipticity
transfer functions have not yet been measured for the assembled cavity. Instead,
we treat this coupling as a commissioning requirement. A full-system
implementation of APE must measure these transfer functions at the chosen
operating point and verify that the resulting angular-jitter noise remains below
the intrinsic birefringence and readout-noise benchmarks used in the present
design study.

The same consideration applies, in a different form, to the QWP configuration:
angular motion of the wave plates can rotate their optical axes and mix
polarization quadratures. For this reason, the comparison presented here should
be interpreted as an optical-design and loss-budget comparison, while the final
low-frequency noise floor must be established experimentally with the complete
instrument.

\subsection{Projected sensitivity and comparison with existing searches}
\label{subsec:projected_sensitivity_comparison}

\begin{figure}[t]
\centering
\includegraphics[width=0.95\linewidth]{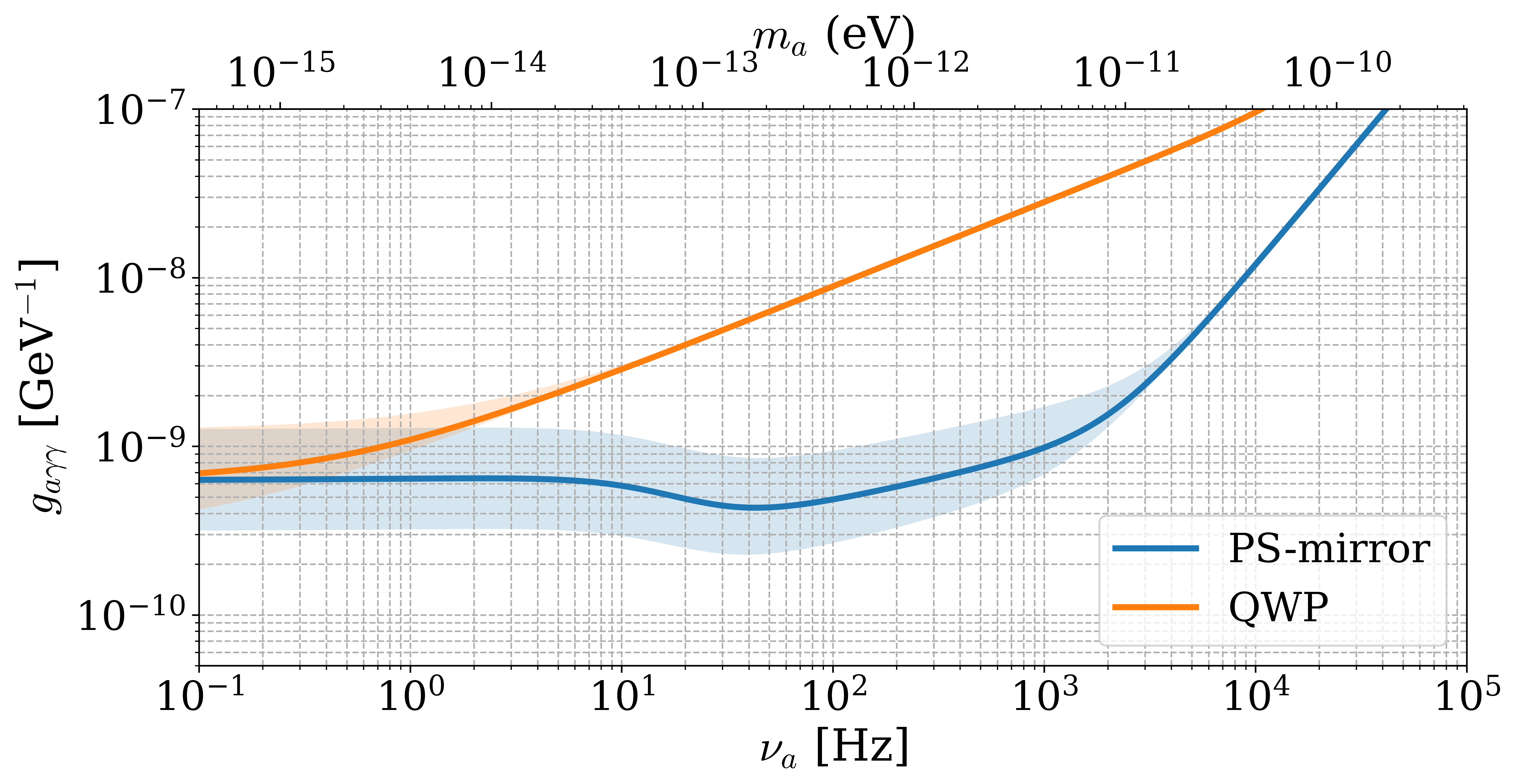}
\caption{\small
Design-level projected sensitivity to the axion--photon coupling
\(g_{a\gamma\gamma}\), including the coating-birefringence benchmark. The solid
curves use the nominal choice \(\chi_{\rm biref}=1\), while the shaded band shows
the effect of varying the coating-birefringence scaling over
\(\chi_{\rm biref}=0.5\text{--}2\). The curves assume the cavity parameters of
Table~\ref{tab:cavity_specs}, input power \(P_0=200~\mathrm{mW}\), local
dark-matter density \(\rho_{\rm DM}=0.4~\mathrm{GeV}/\mathrm{cm}^{3}\), and
coherent integration time \(t_{\rm int}=t_{\rm coh}=10^6/\nu_a\). The orange
curve corresponds to the QWP cavity and the blue curve to the phase-shifting
mirror cavity. These curves are projections under the stated loss, readout, and
coating-birefringence assumptions, not measured exclusion limits.
}
\label{fig:sensitivity_biref}
\end{figure}

Combining the cavity response, heterodyne readout noise, and the intrinsic-birefringence benchmark gives the final design-level
sensitivity projection shown in Fig.~\ref{fig:sensitivity_biref}. At low
frequencies the projection is limited by the assumed intrinsic-birefringence
benchmark, while at higher frequencies it is limited by the cavity transfer
function and the finite coherent integration time. Since the birefringence
spectrum has not yet been measured for the assembled APE cavity, the curves
should be interpreted as projected reach under the stated noise assumptions, not
as measured exclusion limits.


The projected reach in Fig.~\ref{fig:sensitivity_biref} can be compared with
recent optical axion-polarimetry searches. LIDA reported a peak sensitivity of
\(1.51\times10^{-10}~\mathrm{GeV}^{-1}\) at
\(m_a\simeq1.97\text{--}2.01~\mathrm{neV}\), DANCE Act-1 set a
95\% C.L. bound \(g_{a\gamma\gamma}\lesssim
8\times10^{-4}~\mathrm{GeV}^{-1}\) in the range
\(10^{-14}\text{--}10^{-13}~\mathrm{eV}\), and ADBC demonstrated a tunable
birefringent-cavity search with an average constraint
\(g_{a\gamma\gamma}\leq1.9\times10^{-8}~\mathrm{GeV}^{-1}\) in several
windows near \(40\text{--}60~\mathrm{neV}\)
\cite{Heinze:2023nfb,Oshima:2023csb,Pandey2024ADBC}. These results should not
be compared as a simple ranking, since the experiments use different cavity
geometries, readout observables, integration times, and confidence conventions.
Rather, they illustrate the breadth of the emerging optical-polarimetry program.
The added value of APE is to test a low-loss reflective architecture for coherent
polarization buildup: the phase-shifting mirrors replace transmissive
intracavity QWPs while preserving the polarization transformation required for
resonant enhancement of the axion-induced orthogonal field. In the low-mass
short-baseline limit, \(\omega_a L/c\ll1\), the single-pass axion-induced
rotation scales approximately linearly with the optical length. Thus, for fixed
equivalent rotation noise and cavity response, the coupling reach scales roughly
as \(g_{a\gamma\gamma}^{\rm min}\propto 1/L\). The present
\(1.5~\mathrm{m}\) APE cavity should therefore be viewed as a compact
demonstrator of a scalable polarimetric architecture, complementary to existing
cavity-polarimetry searches, to longer-baseline interferometric proposals such as
KAGRA and gravitational-wave detectors, and to non-polarimetric searches such as
ALPS~II \cite{Michimura:2025kod,Michimura:2021hwr,Gottel:2024cfj,LVK:2025dm,ALPSII:2025eri,Spector:2026eys}.

\section{Conclusion}

We have presented the Axion Polarimetric Experiment (APE), a cavity-enhanced polarimeter designed
to search for ultralight axion and axion-like-particle dark matter through the polarization
rotation induced in linearly polarized laser light. The experiment is implemented in two stages:
a QWP-based cavity that establishes the vacuum polarimetric readout and a folded low-loss cavity
based on dielectric phase-shifting mirrors that provides the scalable architecture for the full
axion search.

The central result of this work is that the reflective phase-shifting-mirror design reproduces
the required quarter-wave polarization transformation without transmissive intracavity optics. The
fabricated mirrors realize the target differential reflection phase near the operating angle, and
their measured transmission supports a substantially larger signal-mode finesse than the
QWP-based alternative. When these measured optical properties are propagated through the cavity
and readout model, the resulting configuration yields improved projected sensitivity to the
axion--photon coupling in the low-mass regime. The angle dependence of the phase-shifting mirrors
also provides a practical method for tuning the signal-mode resonance relative to the locked pump.

Our sensitivity estimate combines the cavity transfer function, the PEM-based heterodyne
implementation, the photon-counting benchmark, and an empirical low-frequency noise model
anchored to birefringence measurements in related polarimetric systems. Within this framework,
APE can approach the shot-noise floor over a substantial part of the accessible band, while
intrinsic birefringence and angular-jitter coupling remain the dominant low-frequency
uncertainties to be established experimentally.

Taken together, these results identify reflective phase-shifting cavities as a practical route
toward cavity-enhanced axion polarimetry with substantially improved projected sensitivity
relative to transmissive QWP-based cavities. The next steps are full-system operation, direct
measurement of the intrinsic birefringence noise of the assembled apparatus, measurement of the
yaw- and pitch-to-ellipticity transfer functions of the PSMs, and an end-to-end demonstration of
polarization buildup in a cavity employing the PSM pair. This final test can be performed by
locking the \(p\)-polarized carrier and injecting a calibrated weak \(s\)-polarized field, or an
equivalent polarization modulation, to verify the expected resonant enhancement of the
orthogonal field.
\begin{acknowledgments}
We acknowledge support from the Cluster of Excellence QuantumFrontiers
(EXC 2123/2, DFG Project ID 390837967). This article is based upon work from
COST Action COSMIC WISPers CA21106, supported by COST (European Cooperation in
Science and Technology).
\end{acknowledgments}

\appendix

\section{Multilayer dielectric mirrors}
\label{app:coating}

\subsection{Layer geometry, phase thickness, and wave admittance}
\label{app:coating_geometry}

A dielectric multilayer mirror consists of alternating high- and low-index layers,
\((n_H,n_L)\), deposited between an incident medium of refractive index \(n_0\) and a
substrate of refractive index \(n_s\).

For a layer \(j\) with refractive index \(n_j\) and physical thickness \(d_j\), the
propagation angle inside the layer follows from Snell's law~\cite{YehOWLM},
\begin{equation}
\theta_j
=
\arcsin\!\left(\frac{n_0}{n_j}\sin\theta_0\right),
\label{eq:snell_layer_angle}
\end{equation}
where \(\theta_0\) is the angle of incidence in the incident medium. The corresponding
phase thickness is
\begin{equation}
\delta_j
=
\frac{2\pi}{\lambda}\,n_j d_j\cos\theta_j .
\label{eq:layer_phase_thickness}
\end{equation}

For nonmagnetic materials, \(\mu=\mu_0\), the polarization dependence enters through
the wave admittance \(Y\), defined as the ratio of the transverse magnetic field to the
tangential electric field. For the two linear polarizations one has~\cite{YehOWLM}
\begin{align}
Y_j^{(s)}
&=
\frac{n_j\cos\theta_j}{Z_0},
\label{eq:admittance_s}
\\
Y_j^{(p)}
&=
\frac{n_j}{Z_0\cos\theta_j},
\label{eq:admittance_p}
\end{align}
where \(Z_0=\sqrt{\mu_0/\varepsilon_0}\) is the free-space impedance. Because
\(Y_j^{(s)}\neq Y_j^{(p)}\) at oblique incidence, the multilayer reflectance and
reflection phase are generally different for \(s\)- and \(p\)-polarized light.

\subsection{Characteristic-matrix method}
\label{app:transfer_matrix}

For each polarization \(q\in\{s,p\}\), a single layer is represented by the
characteristic matrix
\begin{equation}
\mathbf{M}_j^{(q)}
=
\begin{pmatrix}
\cos\delta_j
&
i\sin\delta_j/Y_j^{(q)}
\\
iY_j^{(q)}\sin\delta_j
&
\cos\delta_j
\end{pmatrix},
\label{eq:single_layer_matrix}
\end{equation}
which relates the tangential fields on the two sides of the layer~\cite{YehOWLM}.
For a stack of \(N\) layers, numbered from the incident medium toward the substrate,
the total characteristic matrix is the ordered product
\begin{equation}
\mathbf{M}^{(q)}
=
\mathbf{M}_1^{(q)}
\mathbf{M}_2^{(q)}
\cdots
\mathbf{M}_N^{(q)}
=
\begin{pmatrix}
M_{11}^{(q)} & M_{12}^{(q)} \\
M_{21}^{(q)} & M_{22}^{(q)}
\end{pmatrix}.
\label{eq:stack_matrix}
\end{equation}

Let \(Y_0^{(q)}\) and \(Y_s^{(q)}\) denote the admittances of the incident medium
and the substrate, respectively. The input admittance seen from the incident side of
the coating is
\begin{equation}
Y_{\rm in}^{(q)}
=
\frac{
M_{21}^{(q)}+M_{22}^{(q)}Y_s^{(q)}
}{
M_{11}^{(q)}+M_{12}^{(q)}Y_s^{(q)}
}.
\label{eq:input_admittance}
\end{equation}
The corresponding complex reflection coefficient is
\begin{equation}
r_q
=
\frac{Y_0^{(q)}-Y_{\rm in}^{(q)}}
{Y_0^{(q)}+Y_{\rm in}^{(q)}} .
\label{eq:complex_reflection_coefficient}
\end{equation}
From this we obtain the power reflectance and reflection phase,
\begin{equation}
R_q=|r_q|^2,
\qquad
\phi_q=\arg(r_q),
\qquad
\Delta\phi\equiv\phi_s-\phi_p .
\label{eq:reflectance_phase_definitions}
\end{equation}

This formalism is applied at the operating wavelength and angle of incidence to
optimize the layer thicknesses of the phase-shifting mirrors. The design targets are
simultaneously high reflectance for both polarizations and a differential reflection
phase \(\Delta\phi\simeq\pi/2\) in the local \(s/p\) basis.

\section{Ellipsometric characterization of the phase-shifting mirror}
\label{app:ellipsometry}

To determine the differential reflection phase of the fabricated phase-shifting mirror,
we used a PEM-based ellipsometric setup. A QWP--HWP pair prepares a linear polarization
state at \(45^\circ\) with respect to the local \(s/p\) basis, and a PEM imposes a
known time-dependent ellipticity,
\begin{equation}
\eta(t)=\eta_0\cos(\Omega t),
\label{eq:pem_ellipsometry_eta}
\end{equation}
with modulation depth \(\eta_0\) and angular frequency \(\Omega\). A QWP placed after
the mirror compensates the nominal quarter-wave retardance of the reflected field so
that the beam can be extinguished by the analyzer when the mirror phase satisfies
\(\Delta\phi\simeq\pi/2\).

\begin{figure}[ht]
    \centering
    \includegraphics[width=0.9\linewidth]{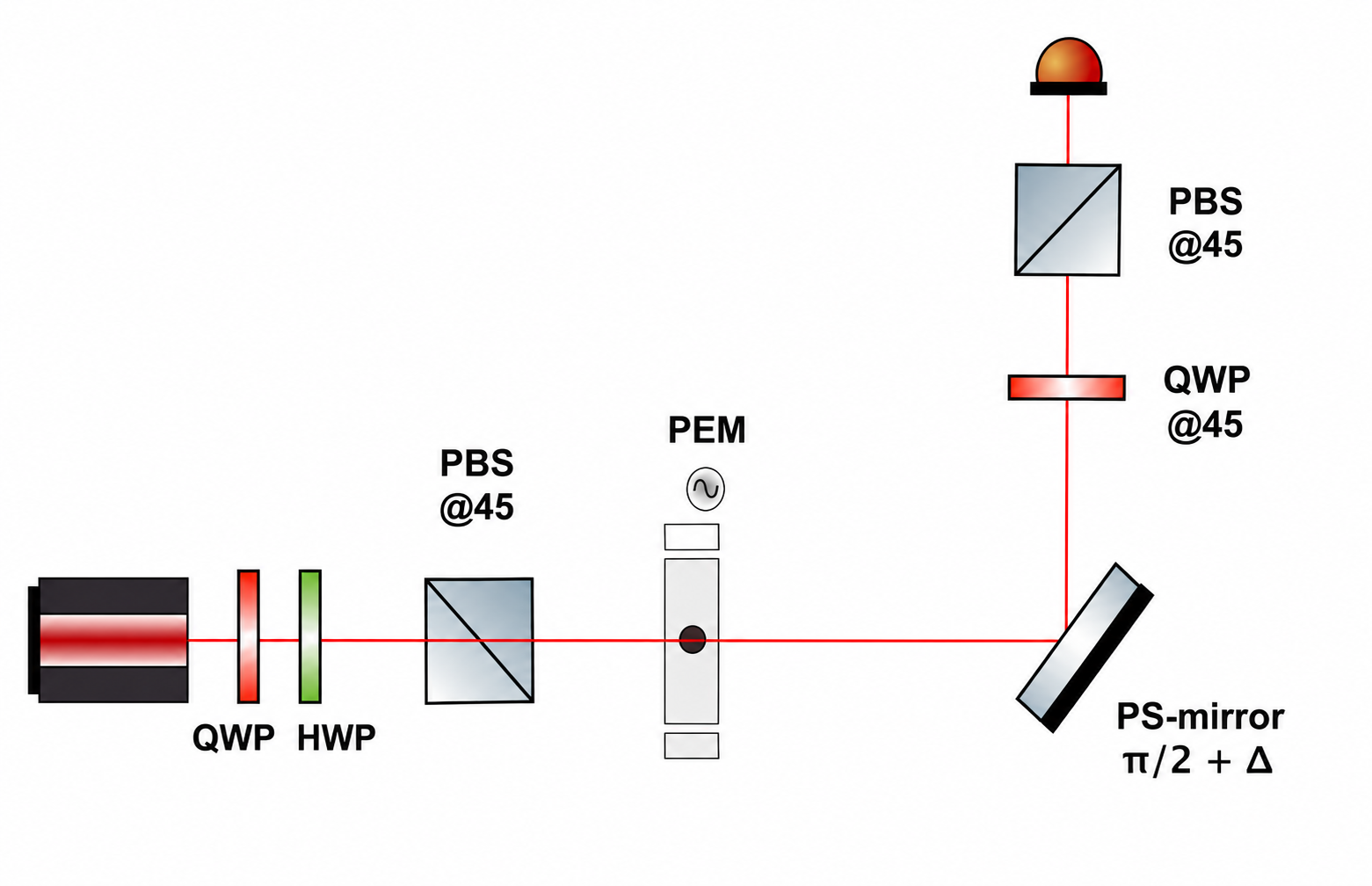}
    \caption{\small
    Ellipsometric setup used to characterize the PS-mirror.  
    A QWP--HWP pair prepares linear polarization at $45^\circ$, which is modulated by the PEM, then reflected from the PS-mirror.  
    Because the PS-mirror introduces a $\pi/2$ phase shift between $s$ and $p$ near $45^\circ$ incidence, a QWP is placed in front of the mirror to convert the reflected field back to a linear state that can be extinguished by the analyzer (PBS @45).  
    The harmonic content of the detected power is used to infer the mirror’s phase error and the PEM modulation depth.}
    \label{fig:psmirror_ellipsometer}
\end{figure}

Writing the retardance error as
\begin{equation}
\Delta
\equiv
\Delta\phi-\frac{\pi}{2},
\label{eq:retardance_error}
\end{equation}
the extinction-port power can be expanded as
\begin{equation}
P_{\rm ext}(t)
=
P_0
\left[
\sigma^2
+
\frac{\eta_0^2}{2}
+
2\Delta\eta_0\cos(\Omega t)
+
\frac{\eta_0^2}{2}\cos(2\Omega t)
\right],
\label{eq:extinction_power_expansion}
\end{equation}
where \(P_0\) is the input power and \(\sigma\) is the extinction ratio of the
polarimeter. The first-harmonic term at \(\Omega\) is proportional to the retardance
error \(\Delta\) and therefore vanishes at the operating point
\(\Delta\phi=\pi/2\). By contrast, the second-harmonic component,
\begin{equation}
P_{2\Omega}(t)
=
\frac{1}{2}P_0\eta_0^2\cos(2\Omega t),
\label{eq:pem_second_harmonic}
\end{equation}
depends only on the PEM modulation depth and is used to calibrate \(\eta_0\) in situ.

In the experiment, the mirror was mounted on a precision rotation stage with
\(5~\mu{\rm rad}\) angular resolution. By scanning the angle of incidence and recording
the corresponding first-harmonic signal, we extracted the retardance error \(\Delta\)
as a function of angle. The resulting angular dependence was then compared with the
transfer-matrix prediction for the coating design, allowing us to identify the operating
angle at which the differential reflection phase satisfies \(\Delta\phi\simeq\pi/2\).

\section{Significance of vacuum operation}
\label{app:vacuum}

Operation in vacuum is required to realize the low-loss and low-noise conditions
assumed in the cavity sensitivity model. The experiment is designed to operate at
pressures below \(10^{-7}~\mathrm{mbar}\). At this pressure, residual-gas loss is
negligible compared with the ppm-level mirror and coating losses.

As a rough estimate, the Rayleigh volume-scattering coefficient of standard air
near \(1064~\mathrm{nm}\) is of order \(10^{-6}~\mathrm{m}^{-1}\)
\cite{Bucholtz:95}. Since the molecular density scales approximately linearly
with pressure, operation at \(10^{-7}~\mathrm{mbar}\) suppresses this coefficient
by a factor
\begin{equation}
\frac{p}{p_{\rm atm}}
\simeq
\frac{10^{-7}~\mathrm{mbar}}{1013~\mathrm{mbar}}
\simeq
10^{-10}.
\end{equation}
The corresponding round-trip Rayleigh-scattering loss over a few-metre optical
path is therefore below \(10^{-15}\). This is many orders of magnitude below the
ppm-level mirror-transmission and coating-loss budget. The finesse is therefore set by the usual round-trip power-loss budget,
\begin{equation}
\mathcal{F}
\simeq
\frac{2\pi}{T_1+T_2+l_{\mathrm{int,rt}}},
\label{eq:finesse_vacuum}
\end{equation}
with the notation defined in Sec.~\ref{section-sensitivity}.

Vacuum operation also suppresses refractive-index fluctuations. The refractivity
of air at optical wavelengths is \(n-1\simeq 2.7\times10^{-4}\) at atmospheric
pressure and scales approximately linearly with gas density
\cite{Ciddor:96}. At \(p<10^{-7}~\mathrm{mbar}\), this gives
\begin{equation}
n-1
\simeq
2.7\times10^{-4}
\frac{p}{p_{\rm atm}}
\lesssim
3\times10^{-14}.
\end{equation}
For a round-trip length of order \(L_{\rm rt}\simeq3~\mathrm{m}\) at
\(\lambda=1064~\mathrm{nm}\), the corresponding static single-round-trip gas
phase is
\begin{equation}
\phi_{\rm gas}
\simeq
\frac{2\pi}{\lambda}(n-1)L_{\rm rt}
\lesssim
5\times10^{-7}~\mathrm{rad}.
\end{equation}

A resonant cavity can enhance intracavity phase or optical-path fluctuations by a
factor of order \(\mathcal{F}/\pi\). However, only fluctuations of
\(\phi_{\rm gas}\), not the static phase itself, contribute to noise. Moreover,
ordinary residual-gas refractivity is isotropic and therefore mainly produces a
common-mode cavity phase shift, rather than an \(s/p\) differential ellipticity
signal. Any remaining coupling to the polarimetric channel would require
anisotropy, spatial gradients, or alignment-dependent effects, and is expected to
be small compared with the coating-birefringence and readout-noise terms
considered in the sensitivity projections.

For the present experiment, vacuum operation ensures that gas-related loss and
refractive-index noise remain negligible compared with the mirror-loss,
coating-birefringence, and readout-noise terms. Residual polarimetric coupling
from gas-density fluctuations should nevertheless be checked during commissioning
by measuring the cavity response as a function of pressure.

\newpage
\bibliographystyle{unsrt}
\bibliography{my.bib} 
\end{document}